\documentclass[onecolumn,final,letterpaper,12pt,journal]{IEEEtran}

\usepackage{amsmath,amssymb,amsfonts,mathtools}
\usepackage{graphicx}
\usepackage{subfig}
\usepackage{epstopdf}
\usepackage[justification=centering]{caption}
\usepackage{cite}
\usepackage{textcomp,enumitem}
\usepackage{changepage}
\usepackage{color}
\usepackage{stmaryrd}
\usepackage{bm,upgreek}
\usepackage{algorithm,algorithmic}
\usepackage{tikz}
\usetikzlibrary{intersections}
\usetikzlibrary {arrows.meta}
\usetikzlibrary {decorations.pathmorphing}

\newcommand{\Var}{\mathrm{Var}}
\newcommand{\Cov}{\mathrm{Cov}}
\newcommand{\trace}{\mathrm{tr}}

\newcommand{\blkdiag}{\mathrm{block\textrm{-}diag}}
\newcommand{\rank}{\mathrm{rank}}

\newcommand{\MSE}{\rm{MSE}}

\newcommand{\diff}{\mathrm{d}}
\newcommand{\Trans}{\mathrm{T}}
\newcommand{\adj}{\mathrm{Adj}}

\def\MatrixFont{\bf}
\def\VectorFont{\bf}

\newcommand{\mA}{{\MatrixFont A}}
\newcommand{\mB}{{\MatrixFont B}}
\newcommand{\mC}{{\MatrixFont C}}
\newcommand{\mD}{{\MatrixFont D}}

\newcommand{\mF}{{\MatrixFont F}}
\newcommand{\mG}{{\MatrixFont G}}
\newcommand{\mH}{{\MatrixFont H}}
\newcommand{\mI}{{\MatrixFont I}}

\newcommand{\mK}{{\MatrixFont K}}
\newcommand{\mL}{{\MatrixFont L}}
\newcommand{\mM}{{\MatrixFont M}}

\newcommand{\mP}{{\MatrixFont P}}
\newcommand{\mQ}{{\MatrixFont Q}}
\newcommand{\mR}{{\MatrixFont R}}
\newcommand{\mS}{{\MatrixFont S}}

\newcommand{\mU}{{\MatrixFont U}}
\newcommand{\mV}{{\MatrixFont V}}

\newcommand{\va}{{\VectorFont a}}

\newcommand{\vc}{{\VectorFont c}}
\newcommand{\vd}{{\VectorFont d}}

\newcommand{\vu}{{\VectorFont u}}
\newcommand{\vv}{{\VectorFont v}}
\newcommand{\vw}{{\VectorFont w}}
\newcommand{\vx}{{\VectorFont x}}
\newcommand{\vy}{{\VectorFont y}}

\newcommand{\R}{{\mathbb R}}
\newcommand{\sS}{{\mathbb S^+}}
\newcommand{\Z}{{\mathbb Z^+}}
\newcommand{\N}{\mathcal N}
\newcommand{\E}{\mathbb E}
\newcommand{\sP}{{\mathcal P}}

\newcommand{\DK}{\mathrm{DK}}
\newcommand{\HF}{\mathrm{HF}}
\newcommand{\F}{\mathrm F}
\newcommand{\G}{\mathrm G}
\newcommand{\rH}{\mathrm H}
\newcommand{\rK}{\mathrm K}
\newcommand{\Q}{\mathrm Q}
\newcommand{\rR}{\mathrm R}
\newcommand{\I}{\mathcal I}
\newcommand{\B}{\mathrm{CRLB}}
\newcommand{\PB}{\mathrm{PCRLB}}

\newcommand{\Round}{\mathrm{round}}

\def\dx{n_x}
\def\dy{n_y}
\def\dd{n_d}
\def\ns{m}
\def\UMV{\mathrm{umv}}
\def\M{\mathcal M}

\def\vmu{\bm \mu}
\def\vtheta{\bm \theta}
\def\valpha{\bm \alpha}
\def\mSigma{\bm \Sigma}
\def\mLambda{\bm \Lambda}
\def\mGamma{\bm \Gamma}
\def\mDelta{\bm \Delta}
\def\mPhi{\bm \Phi}
\def\mUpsilon{\bm \Upsilon}

\newtheorem{theorem}{Theorem}
\newtheorem{lemma}{Lemma}
\newtheorem{proposition}{Proposition}
\newtheorem{remark}{Remark}
\newtheorem{example}{Example}

\newtheorem{definition}{Definition}
\newtheorem{assumption}{Assumption}
\newtheorem{corollary}{Corollary}

\renewcommand{\IEEEQED}{\IEEEQEDopen}
\IEEEoverridecommandlockouts
\allowdisplaybreaks[3]
\begin{document}


\title{State Estimation with Protecting Exogenous Inputs via Cram\'er-Rao Lower Bound Approach
	\thanks{
		The work was supported by National Natural Science Foundation of China under Grants 62025306, 62433020, 62203045 and T2293770, CAS Project for Young Scientists in Basic Research under Grant YSBR-008, CAS Special Research Assistant Project under Grant E4559305, Postdoctoral Fellowship Program of CPSF under Grant GZB20230813, China Postdoctoral Science Foundation under Grant 2024M763479. The material in this paper was not presented at any conference. \emph{(Corresponding author: Ji-Feng Zhang)}}
	\thanks{Liping Guo and Yanlong Zhao are with the Key Laboratory of Systems and Control, Institute of Systems Science, Academy of Mathematics and Systems Science, Chinese Academy of Sciences, Beijing 100190, China. (e-mail: lipguo@outlook.com; ylzhao@amss.ac.cn)}
	\thanks{Jimin Wang is with the School of Automation and Electrical Engineering, University of Science and Technology Beijing, Beijing 100083, and also with the Key Laboratory of Knowledge Automation for Industrial Processes, Ministry of Education, Beijing
		100083, China. (e-mail: jimwang@ustb.edu.cn)}
	\thanks{Ji-Feng Zhang is with the School of Automation and Electrical Engineering, Zhongyuan University of Technology, Zheng Zhou 450007; and also with the Key Laboratory of Systems and Control, Institute of Systems Science, Academy of Mathematics and Systems Science, Chinese Academy of Sciences, Beijing 100190, China. (e-mail: jif@iss.ac.cn)}}
\author{Liping Guo, \IEEEmembership{Member,~IEEE}, Jimin Wang, \IEEEmembership{Member,~IEEE}, Yanlong Zhao,~\IEEEmembership{Senior Member,~IEEE},
and Ji-Feng Zhang,~\IEEEmembership{Fellow,~IEEE}}

\maketitle
\IEEEpeerreviewmaketitle
\thispagestyle{plain}

\begin{abstract}
This paper addresses the real-time state estimation problem for dynamic systems while protecting exogenous inputs against adversaries, who may be honest-but-curious third parties or external eavesdroppers. 
The Cram\'er-Rao lower bound (CRLB) is employed to constrain the mean square error (MSE) of the adversary's estimate for the exogenous inputs above a specified threshold.
By minimizing the MSE of the state estimate while ensuring a certain privacy level measured by CRLB, the problem is formulated as a constrained optimization. 
To solve the optimization problem, an explicit expression for CRLB is first provided.
As the computational complexity of the CRLB increases with the time step, a low-complexity approach is proposed to make the complexity independent of time.
Then, a relaxation approach is proposed to efficiently solve the optimization problem. 
Finally, a privacy-preserving state estimation algorithm with low complexity is developed,
which also ensures $(\epsilon, \delta)$-differential privacy. Two illustrative examples, including a practical scenario for protecting building occupancy, demonstrate the effectiveness of the proposed algorithm.
\end{abstract}


\section{Introduction}
An increasing array of applications necessitates users to send private data streams to a third party for signal processing and decision-making \cite{LeNy2014,Li2024}. 
Due to privacy concerns, users prefer to send information while protecting their sensitive data. 
As a result, privacy issues have gained increased attention in several research fields, such as mobile robotic systems \cite{Zhang2024}, network systems \cite{Lu2020,Wang2019}, and control systems \cite{Kawano2020,Gao2021,TAN-2023-Cooperative}.  

In recent years, privacy has been widely studied in various fields, e.g., machine learning \cite{tong2024multi,Wei-2024-Gradient,miao2024efficient}, Nash equilibrium \cite{Ye-2022-Differentially,Wang2024}, decentralized optimization \cite{Wang2024c, Lu-2018-Privacy}, and real-time state estimation \cite{LeNy2014, Moradi2022,Weng-2023-Optimal}. In practice,  the state estimates or measurements could be acquired by honest-but-curious third parties or external eavesdroppers, leading to a privacy leakage \cite{Moradi2022,Weng-2023-Optimal,Nekouei2023,Feng2024}. 
For instance, 
in intelligent transportation systems, users are often required to send measurements to a third party for monitoring or control purposes, potentially compromising their privacy \cite{LeNy2014}.
Besides, in environmental monitoring, the building occupancy could be reliably estimated from the CO$_2$ levels \cite{Weng-2023-Optimal,Nekouei2023}. 
Similarly, the thermal dynamics of a building may also result in a privacy leakage of occupancy \cite{Feng2024}. 
Therefore, the privacy preservation problem is an important issue worth studying in real-time state estimation.

Among various privacy techniques, the most researched are encryption \cite{Lu-2018-Privacy,Shang-2022-Linear,Chen-2023-Privacy},  information-theoretic approach \cite{Farokhi-2018-Fisher,Weng-2023-Optimal}, and differential privacy \cite{Dwork2013,Zhang-2021-Privacy}.
Due to its ease of implementation, the differential privacy stands out compared to its competitors. Besides, due to its resilience to post-processing, differential privacy makes reverse engineering of the private datasets difficult, and thus, has been widely adopted to protect privacy 
in stochastic aggregative games \cite{Wang-2022-Differentially}, distributed optimization \cite{Liu-2024-Distributed}, and real-time state estimation \cite{LeNy2014, Dwork2013, LeNy2020}. 
Especially for differentially private state estimation,
the Kalman filtering problem with input and output perturbation mechanisms has been firstly addressed in \cite{LeNy2014}. This work was extended to multi-input multi-output systems, which broadens the applicability to multiple sensors for monitoring an environment \cite{LeNy2018}. Then, a two-stage architecture was proposed in \cite{Degue2023} to overcome the drawback of output perturbation mechanisms. 

Another common approach for privacy preservation is using metrics from information theory to measure private information leakage in response to a query on private datasets \cite{Liu-2017-Information}.
An information-theoretic measure of privacy often relies on
the mutual information, which measures private information leakage by formulating the privacy problem as a generalized rate-distortion problem \cite{Issa-2020-An}.
In addition, mutual information, commonly used to measure the correlation between two random variables, also appears in content related to real-time state estimation problems. 
For instance, 
an information-theoretic framework for the privacy-aware optimal state estimation was proposed in \cite{Weng-2023-Optimal}, where the private dataset was modeled as a first-order Markov process. 
However, as analyzed convincingly in \cite{Farokhi-2019-Ensuring}, most mutual-information-based results lack an intuitive or interpretable bound on the statistics of the estimation error by an adversary. 
To address this limitation, a data-privacy approach was introduced in \cite{He2018} by measuring the adversary's estimation error with the absolute error metric. However, the absolute error metric is generally mathematically challenging to handle.
Consequently, the Fisher information matrix and the Cram\'er-Rao lower bound (CRLB) have been proposed as alternative frameworks  \cite{Farokhi-2018-Fisher,Farokhi-2019-Ensuring,Farokhi-2020-Privacy,Ziemann-2020-Parameter}.  
Although successful in treating control problems \cite{Ziemann-2020-Parameter}, Fisher information matrix does not directly capture the performance of an adversary's estimation accuracy. 
Therefore, the CRLB, based on mean square error (MSE), is more widely adopted in practice \cite{Farokhi-2018-Fisher,Farokhi-2019-Ensuring,Farokhi-2020-Privacy}. 
However, existing studies  \cite{Farokhi-2018-Fisher,Farokhi-2019-Ensuring,Farokhi-2020-Privacy} have focused on protecting parameters that are time-invariant only. In contrast, when addressing the protection of time-varying states in stochastic dynamic systems, the computational complexity continues to increase as time progresses, thereby posing significant challenges. As far as we know, research on CRLB-based privacy preservation is still lacking in real-time state estimation. 

Motivated by the above analysis, 
in this paper, we investigate the real-time state estimation problem with protecting exogenous inputs via CRLB. 
The merit of adopting CRLB lies in that it constrains the MSE of the adversary's estimate for the exogenous inputs above a specified threshold.  
However, there exist some substantial difficulties in studying this problem. 
First, a primary issue is how to ensure privacy level and state estimation accuracy simultaneously? To do so, we face with solving a non-convex constrained optimization problem, which is challenging.
Second, in CRLB-based privacy preservation, calculating CRLB is necessary, but an explicit expression for CRLB is generally difficult to obtain.
Third, the computational efficiency is critically important for real-time state estimation, but the computational complexity of the CRLB tends to become greater over the time step.
Fourth, given that both are based on noise perturbation strategy, is it possible to correlate the proposed CRLB-based privacy preservation with the differential privacy?
These difficulties are solved in this paper and
the main contributions are summarized as follows:
\begin{adjustwidth}{0pt}{}
	\begin{enumerate}[label=\textbullet]
		\item This paper achieves real-time state estimation while protecting exogenous inputs. By employing the CRLB, the MSE of the adversary's estimate for the exogenous inputs is constrained above a specified threshold.
		By minimizing the MSE of state estimate while ensuring a certain privacy level measured by CRLB, a constrained optimization problem is constructed to ensure privacy level and state estimation accuracy simultaneously. Furthermore, a relaxation approach is proposed to efficiently solve the optimization problem.
		\item An explicit expression for the CRLB is provided, laying the foundation for CRLB-based privacy preservation. Furthermore,
		a low-complexity approach for calculating the CRLB is proposed. As a result, the computational complexity is significantly reduced from $\mathcal{O}(k^3)$ to $\mathcal{O}(1)$, which means that the computational complexity of the CRLB is reduced to be time-independent. 
		This development makes valuable sense for real-time state estimation.
		\item A privacy-preserving state estimation algorithm with low complexity is developed. Moreover, the relationship between our proposed CRLB-based privacy preservation and differential privacy is established. Specifically, the proposed algorithm is proven to ensure $(\epsilon, \delta)$-differential privacy. Finally, the effectiveness of the proposed algorithm is demonstrated through two illustrative examples, including a practical scenario for protecting building occupancy. 
	\end{enumerate}
\end{adjustwidth}

The reminder of this paper is organized as follows. Section \ref{sec:problem formulation} formulates the problem.   Section \ref{sec:estimator design} provides the design of the privacy-preserving state estimate and an explicit expression for the CRLB. Section \ref{sec:online state estimation} presents the privacy-preserving state estimation algorithm with low complexity and its relation to differential privacy, followed by two examples in Section \ref{sec:simulation}. Section \ref{sec:conclusion} concludes this paper. 

\textit{Notations.}
Scalars, vectors and matrices are denoted by lowercase letters, bold lowercase letters, and bold capital letters, respectively. 
Scalar $0$, zero vector, and zero matrix are all denoted by $0$ for simplicity.
The set of all $n$-dimensional real vectors and all $n \times m$ real matrices are denoted by $\R^{n}$ and $\R^{n \times m}$, respectively.  
$\Z$ represents the set of positive integers and
$\sS$ represents the set of positive semi-definite matrices.  
For a square matrix $\mA$, $\trace(\mA)$ denotes its trace. $\mA \geq 0$ (or $\mA > 0$) means that $\mA$ is positive semi-definite (or positive definite).
The $\blkdiag(\mA_0, \mA_1, \dots, \mA_k)$ represents the block diagonal matrix with matrices $\mA_0, \mA_1, \dots, \mA_k$ on the principal diagonal. 
For the sequence of square matrices $\mA_0, \mA_1, \dots, \mA_k$ with the same dimension, we define
$\prod_{i = 0}^k \mA_i = \mA_k \mA_{k - 1} \cdots \mA_0$, 
which means that they follow the descending order of the subscript.
$\mI_n$ represents the $n \times n$ identity matrix. 
For a vector $\va$, its Euclidean norm is denoted by $\|\va\|$. Further, $\|\va\|_{\mA}$ denotes its Euclidean norm weighted with $\mA > 0$, i.e., $\sqrt{\va^{\Trans} \mA \va}$. 
$\E[\cdot]$ is the mathematical expectation operator. 
$\N(\vmu, \mSigma)$ denotes the Gaussian distribution with mean $\vmu$ and covariance matrix $\mSigma$. 
Besides, $f(x) = \mathcal{O}(g(x))$ if there exists a positive real number $m$ such that $|f(x)| \leq m g(x)$.

\section{Problem formulation}\label{sec:problem formulation}

Consider the following stochastic time-varying dynamic system:
\begin{equation}\label{eq:system}
	\begin{aligned}
		\vx_k &= \mF_{k-1} \vx_{k-1} + \mG_{k-1} \vd_{k-1} + \vw_{k-1},\\
		\vy_k &= \mH_k \vx_k + \vv_k,
	\end{aligned}
\end{equation}
where $k = 1, 2, \dots$ is the time index; $\vx_k \in \R^{\dx}$, $\vy_k \in \R^{\dy}$, and $\vd_{k-1} \in \R^{\dd}$ are the state, the measurement, and the exogenous input that should be protected at time step $k$, respectively; $\mF_{k-1} \in \R^{\dx \times \dx}$, $\mG_{k-1} \in \R^{\dx \times \dd}$, and $\mH_k \in \R^{\dy \times \dx}$ are known matrices; $\{\vw_k\}$ and $\{\vv_k\}$ are mutually independent Gaussian white noise sequences with zero mean and known covariance matrices $\mQ_k \geq 0$ and $\mR_k > 0$, respectively; $\vx_0 \sim \N(\bar \vx_0, \mP_0)$ is the initial state independent of the noise sequences. 
We assume that there is no available prior information about the exogenous input $\vd_{k-1}$. 
Under this case, $\vd_{k-1}$ is generally modeled as a deterministic, time-varying, but unknown (or uncertain) quantity (see, e.g., \cite{kitanidis1987unbiased,darouach1997unbiased}). 

\begin{assumption}\label{assum:full rank}
	$\rank(\mH_k \mG_{k-1}) = \rank(\mG_{k-1}) = \dd$, for all $k$.
\end{assumption}

\begin{remark}
	Assumption \ref{assum:full rank} is commonly used in existing literature,  e.g., \cite{kitanidis1987unbiased,darouach1997unbiased}. 
	Note that Assumption \ref{assum:full rank} indicates that $\dx \geq \dd$ and $\dy \geq \dd$.
\end{remark}

\begin{figure*}[htbp]
	\centering
	\begin{tikzpicture}[scale = 1.0,>= Stealth]
		\draw[->, line width= 0.8] (0, 0) -- node[anchor = south]{$\vd_{k-1}$} (2.2, 0) coordinate (a);
		\draw (a)+(-1.1, -0.5) node[anchor = north, text width = 1.2cm, text centered]{exogenous input};
		\draw[line width = 0.8] (a)++(0, -0.5) rectangle node{system} +(2, 1) coordinate (a);
		\draw[->, line width= 0.8] (a)++(0, -0.5) -- node[anchor = south]{$\vy_k$}  +(2.2, 0) coordinate (a);
		\draw (a)+(-1.1, -0.7) node[anchor = north]{measurement};
		\draw[line width = 0.8] (a)++(0, -0.5) rectangle node{estimator} +(2, 1) coordinate (a);
		\draw [->, line width = 0.8] (a)++(0, -0.5) -- node[anchor = south]{$\hat \vx_k$}  +(2.2, 0) coordinate (a);
		\draw (a)+(-1.1, -0.55) node[anchor = north, text width = 1.8cm, text centered]{state estimate};
		\draw[line width = 0.8] (a)++(0, -0.5) rectangle node{adversary} +(2, 1) coordinate (a);
		\draw[->, line width= 0.8] (a)++(0, -0.5) -- node[anchor = south]{$\hat \vd_{k-1}$}  +(2.2, 0) coordinate (a);
		\draw (a)+(-1.1, -0.5) node[anchor = north, text width = 2.6cm, text centered]{exogenous input estimate};
	\end{tikzpicture}
	\captionof{figure}{The privacy-preserving state estimation setup.}
	\label{fig:privacy-aware state estimation setup}
\end{figure*}

Let $\hat \vx_k$ be the estimate of the state $\vx_k$ using the measurements $\vy_0, \vy_1, \dots, \vy_k$. 
In our setup, $\vd_k$ drives the system, and the state estimate $\hat \vx_k$ should be transmitted to a third party for signal processing or decision-making purposes. In this process, an adversary might have the ability to acquire $\hat \vx_k$ and use it to infer $\vd_{k-1}$, as illustrated in Fig.~\ref{fig:privacy-aware state estimation setup}. 
In this work, the adversary could be an honest-but-curious third party or an external eavesdropper, and is assumed to have the following two abilities:
\begin{itemize}
	\item It can acquire the system matrices $\mF_{k - 1}$, $\mG_{k - 1}$,  $\mQ_{k-1}$, $\mH_k$ and $\mR_k$;
	\item It can store and utilize the $\ns$ ($\ns \in \Z $ and $\ns \geq 2$) state estimates $\hat \vx_{k - \ns + 1}, \hat \vx_{k - \ns + 2}, \dots, \hat \vx_k$ to infer $\vd_{k - 1}$.
\end{itemize}
It should be noted that the second ability makes sense in practice since the adversary cannot store infinite data. 

In this paper, we aim to design a state estimate $\hat \vx_k$ that minimizes the MSE of $\vx_k$ on the premise of protecting $\vd_{k - 1}$.
It is worth emphasizing that we focus on protecting the latest exogenous input only at each time step. In another word, at time step $k$, our goal is to protect $\vd_{k-1}$. This makes sense in many practical scenarios. For instance, in the motivating example given by Example \ref{subsec:motivate:exam} later, the adversary is interested in the current  building occupancy rather than the past. 
Before further discussion, we first briefly review the optimal state estimate in the minimum MSE sense for the system \eqref{eq:system} without privacy consideration, which is the unbiased minimum-variance state estimate proposed in \cite{kitanidis1987unbiased}. 

\subsection{Unbiased minimum-variance state estimate}


Let $\hat \vx_{k - 1}^{\UMV}$ and $\hat \mS_{k -  1}^{\UMV}$, respectively, be the unbiased minimum-variance state estimate and its error covariance matrix at time step $k - 1$. Let $\hat \vx_k^-$ be the estimate of the state $\vx_k$ using the measurements $\vy_0, \vy_1, \dots, \vy_{k - 1}$, and $\hat \mS_k^-$ be the error covariance matrix of $\hat \vx_k^-$. Then, the one-step prediction is given as
\begin{align}
	\hat \vx_k^- &= \mF_{k - 1} \hat \vx_{k - 1}^{\UMV}, \label{eq:UMV:pre:mean}\\
	\hat \mS_k^- &= \mF_{k - 1} \hat \mS_{k - 1}^{\UMV} \mF_{k - 1}^{\Trans} + \mQ_{k - 1}.\label{eq:UMV:pre:cov}
\end{align}
Once receiving the measurement $\vy_k$, the unbiased minimum-variance state estimate and its error covariance matrix at time step $k$ are given as
\begin{align}
	\hat \vx_k^{\UMV} &= \hat \vx_k^- + \mK_k (\vy_k - \mH_k \hat \vx_k^-), \label{eq:UMV estimate}\\
	\hat \mS_k^{\UMV} &= \hat \mS_k^- - \hat \mS_k^-  \mH_k^{\Trans} \mC_k^{-1} \mH_k \hat \mS_k^- 
	 + (\mG_{k - 1} - \hat \mS_k^- \mH_k^{\Trans} \mC_k^{-1} \mH_k \mG_{k - 1})  
	 (\mG_{k - 1}^{\Trans} \mH_k^{\Trans} \mC_k^{-1} \mH_k \mG_{k - 1})^{-1} \notag\\
	&\quad \cdot (\mG_{k - 1} - \hat \mS_k^- \mH_k^{\Trans} \mC_k^{-1} \mH_k \mG_{k - 1})^{\Trans}, \label{eq:UMV:update:cov}
\end{align}
where
\begin{align*}
	\mK_k &= \hat \mS_k^- \mH_k^{\Trans} \mC_k^{-1} + (\mG_{k - 1} - \hat \mS_k^- \mH_k^{\Trans} \mC_k^{-1} \mH_k \mG_{k - 1}) (\mG_{k - 1}^{\Trans} \mH_k^{\Trans} \mC_k^{-1} \mH_k \mG_{k - 1})^{-1} \mG_{k - 1}^{\Trans} \mH_k^{\Trans} \mC_k^{-1}, \\
	\mC_k &= \mH_k \hat \mS_k^- \mH_k^{\Trans} + \mR_k.
\end{align*}
Despite achieving good state estimation accuracy, the unbiased minimum-variance state estimate given by \eqref{eq:UMV estimate} cannot be transmitted directly as it may cause a privacy leakage of $\vd_{k-1}$, as analyzed below.

\subsection{Privacy issue}

This section presents an illustrative example to demonstrate why the unbiased minimum-variance state estimate given by \eqref{eq:UMV estimate} may cause a privacy leakage of $\vd_{k-1}$. 
As the inference approach employed by the adversary is neither unique nor the main concern of this paper, we next construct a straightforward but suboptimal estimate $\hat \vd_{k - 1}$ serving as an illustrative example. Specifically,
from the state transition equation in \eqref{eq:system}, we have 
$\mG_{k-1} \vd_{k-1} = \vx_k - \mF_{k-1} \vx_{k-1} - \vw_{k-1}$.
From Assumption \ref{assum:full rank},
by multiplying both sides of the above equation by $\mG_{k - 1}^{\Trans}$ and plugging in $\hat \vx_k$, $\hat \vx_{k-1}$ and $\hat \vw_{k - 1} = 0$, we can construct the following estimate of $\vd_{k-1}$:
\begin{align}\label{eq:input infer}
	\hat \vd_{k - 1} = (\mG_{k - 1}^{\Trans} \mG_{k - 1})^{-1} \mG_{k - 1}^{\Trans} ( \hat \vx_k -  \mF_{k - 1} \hat \vx_{k - 1}).
\end{align} 
We know from \eqref{eq:input infer} that the adversary could estimate $\vd_{k - 1}$ by using the state estimates $\hat \vx_k$ and $\hat \vx_{k - 1}$.  
It should be noted that $\hat \vd_{k - 1}$ given by \eqref{eq:input infer} is a suboptimal estimate because only two state estimates, $\hat \vx_k$ and $\hat \vx_{k - 1}$, are used. 


To demonstrate the necessity of protecting $\vd_{k-1}$, several practical examples from the literature can be considered. For instance, in building automation systems, it is crucial to prevent the inference of occupancy levels from observable measurements such as CO$_2$ concentrations or temperature dynamics \cite{Nekouei-2022-Optimal,Weng-2023-Optimal}. Similarly, in smart grid applications, protecting a user's electricity consumption data from being accurately inferred by adversaries represents another key scenario where privacy is essential \cite{Li-2018-Information-Theoretic}. 
These examples highlight the broad relevance of protecting $\vd_{k-1}$
across different domains.
The following example is provided to illustrate the privacy leakage of $\vd_{k-1}$ caused by the unbiased minimum-variance state estimate given by \eqref{eq:UMV estimate} in a practical scenario.

\begin{example}[Building occupancy]\label{subsec:motivate:exam}

Similar to \cite{Weng-2023-Optimal,Nekouei2023}, we consider the evolution of CO$_2$ in a building, which can be modeled by the  dynamic equation
$x_k = a x_{k - 1} + b d_{k - 1} + w_{k - 1},
$
where $x_k$ is the level of CO$_2$ at time step $k$, $d_{k - 1}$ is the occupancy of the building, i.e., the number of people in the building, $w_{k - 1}$ is the process noise, and $a, b \in \R$ are the parameters. The measurement at time step $k$ is collected by a CO$_2$ sensor with a measurement equation given by
$
	y_k = x_k + v_k,
$
where $v_k$ is the measurement noise.

The occupancy of the building is sensitive and highly private information that could be reliably estimated from the CO$_2$ levels (see, e.g., \cite{Weng-2023-Optimal}). To demonstrate this fact in our setup, we simulate the CO$_2$ evolution by taking $a = 0.75$, $b = 1.75$, and model $\{w_k\}$ and $\{v_k\}$ as Gaussian white noise sequences with variances $0.1$ and $0.05$, respectively. The initial state is distributed from the Gaussian distribution, with a mean and variance of $0.01$. The real privacy state is given as $d_{k - 1} = \Round(0.5 \cos(k) + 5)$, where $\Round(\cdot)$ is the rounding function. 

%

\begin{figure}[htbp]
	\centering
	\subfloat[CO$_2$ level and its estimate by the unbiased minimum-variance state estimate.]{
		\includegraphics[width=0.6\textwidth]{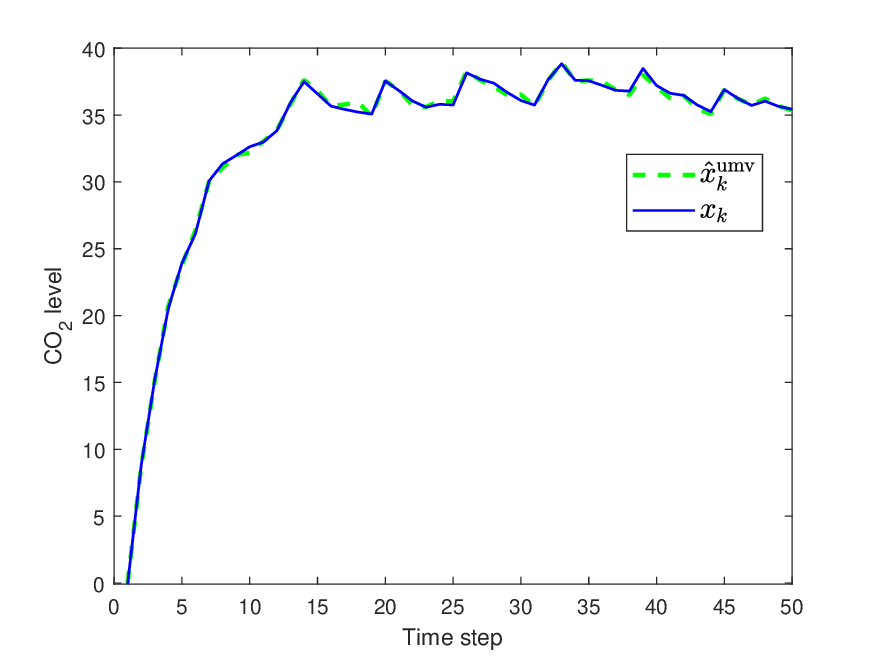}
		\label{fig:state_estimate}
	}
	\hfill
	\subfloat[Occupancy and its estimate by the adversary using \eqref{eq:input infer}.]{
		\includegraphics[width=0.6\textwidth]{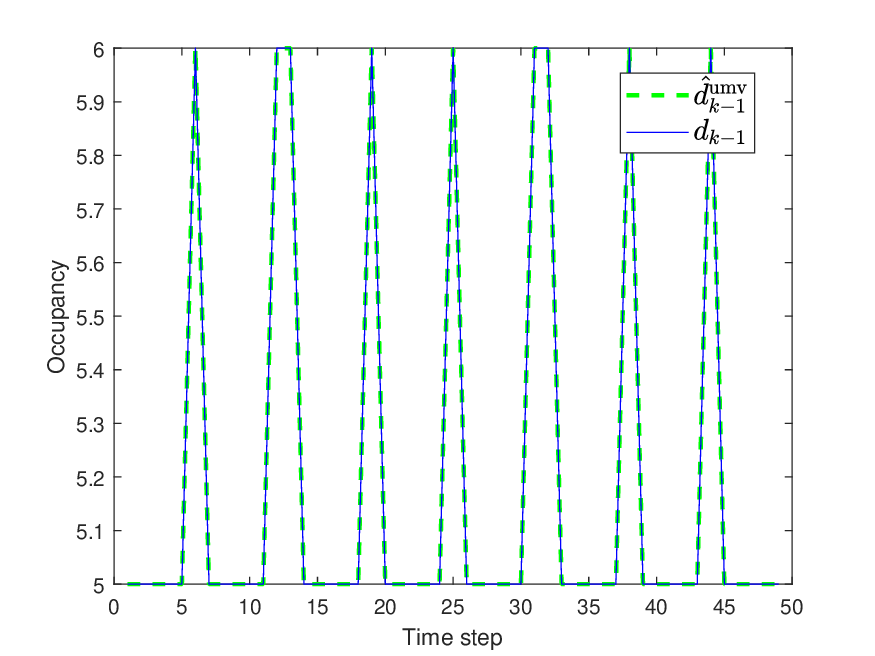}
		\label{subfig:occupancy}
	}
	\caption{CO$_2$ level, occupancy and their estimates.}
	\label{fig:1111}
\end{figure}

Fig.~\ref{fig:state_estimate} depicts the trajectory of the CO$_2$ level in the building and the estimated trajectory by the unbiased minimum-variance state estimate given by \eqref{eq:UMV estimate}. 
Fig.~\ref{subfig:occupancy} illustrates the trajectory of the occupancy and the estimated trajectory by the adversary using \eqref{eq:input infer}, where the state estimates used in \eqref{eq:input infer} are $\hat \vx_k^{\UMV}$ and $\hat \vx_{k-1}^{\UMV}$.  
The plots reveal that the CO$_2$ level and the occupancy are well estimated, suggesting that transmitting the CO$_2$ estimates to the third party could result in a privacy loss of the occupancy.
\end{example}

Based on the above analysis, we know that the unbiased minimum-variance state estimate cannot be transmitted directly, and thus, a privacy-preserving state estimate in the minimum MSE sense should be designed.

\section{Design of the privacy-preserving state estimate via CRLB}\label{sec:estimator design}

In this section, we provide the privacy-preserving state estimation in the minimum MSE sense via CRLB. 
Inspired by the noise perturbation strategy, we consider perturbing the unbiased minimum-variance state estimate with a random noise. 

\subsection{Perturbed noise approach}

We design the privacy-preserving state estimate $\hat \vx_k$ by perturbing $\hat \vx_k^{\UMV}$ with a zero-mean Gaussian noise as follows:
\begin{align}\label{eq:pp state estimator}
	\hat \vx_{k} = \hat \vx_k^{\UMV} + \valpha_k,
\end{align}
where $\valpha_k  \sim \N(0, \mSigma_k)$ is independent of $\hat \vx_k^{\UMV}$, and the covariance matrix $\mSigma_k$ is to be determined. The determination of $\mSigma_k$ is the main concern of our noise perturbation strategy since it determines not only the estimation accuracy of $\hat \vx_{k}$, but also the privacy level of $\vd_{k-1}$. 
To measure the privacy level, we employ the CRLB,  
as specified by the following lemma.

\begin{lemma} [Cram\'er-Rao lower bound, \cite{Shao-Mathematical-2003}] \label{thm:CRLB}
	Let $X = \{\vx_1, \dots, \vx_{\ns}\}$ be a sample from $P \in \sP = \{P_{\vtheta}: \vtheta \in \Theta\}$, where $\Theta$ is an open set in $\R^m$. Suppose that $T(X)$ is an estimator with $\E [T(X)] = g(\vtheta)$ being a differential function of $\vtheta$, $P_{\vtheta}$ has a probability density function $p_{\vtheta}$ with respect to a measure $\nu$ for all $\vtheta \in \Theta$, and $p_{\vtheta}$ is differential as a function of $\vtheta$ and satisfies
$
		\frac{\partial}{\partial \vtheta} \int h(\vx) p_{\vtheta}(\vx) = \int h(\vx) \frac{\partial}{\partial \vtheta} p_{\vtheta}(\vx) \diff \nu, \ \vtheta \in \Theta,
$
	for $h(\vx) \equiv 1$ and $h(\vx) = T(\vx)$. Then,
	\begin{align}\label{ineq:CRLB}
		\Var(T(X)) \geq \bigg(\frac{\partial}{\partial \vtheta} g(\vtheta)\bigg)^{\Trans} \big(\I(\vtheta)\big)^{-1} \frac{\partial}{\partial \vtheta} g(\vtheta),
	\end{align}        
	where 
$\I(\vtheta) = \E[\frac{\partial}{\partial \vtheta} \log p_{\vtheta}(X) (\frac{\partial}{\partial \vtheta} \log p_{\vtheta}(X))^{\Trans}]$         is the Fisher information matrix and assumed to be positive definite for any $\vtheta \in \Theta$.                        
\end{lemma}

In this work, $X = \{\hat \vx_{k - \ns + 1}, \hat \vx_{k - \ns + 2}, \dots, \hat \vx_k\}$, $T(X) = \hat \vd_{k-1}$, $\vtheta = [\vd_0^{\Trans}, \vd_1^{\Trans}, \dots, \vd_{k-1}^{\Trans}]^{\Trans}$, $g(\vtheta) = \vd_{k-1}$.
We know from \eqref{ineq:CRLB} that CRLB provides a lower bound for the 
MSE matrices
of all the unbiased estimators, i.e., 
provides an intuitive quantified metric for the performance of all the unbiased estimators. 
In this paper, the merit of adopting CRLB lies in that it constrains the MSE of the adversary's estimate for the $\vd_{k - 1}$ above a specified threshold. 
We next provide a way of determining the covariance matrix $\mSigma_k$ through a constrained optimization with privacy constraint measured by CRLB.

\subsection{Constrained optimization subject to privacy}

We determine the covariance matrix $\mSigma_k$ by minimizing the MSE of \eqref{eq:pp state estimator} as follows:
\begin{align}\label{eq:MSE of optimal estimator}
	\trace \E\Big[(\hat \vx_k - \vx_k) (\hat \vx_k - \vx_k)^{\Trans}\Big] 
	= \trace \E\Big[(\hat \vx_k^{\UMV} - \vx_k) (\hat \vx_k^{\UMV} - \vx_k)^{\Trans}\Big] + \E\Big[\valpha_k \valpha_k^{\Trans}\Big] 
	= \trace(\hat \mS_k^{\UMV} + \mSigma_k),
\end{align}
and by constraining the MSE of the adversary's estimate for the $\vd_{k - 1}$ above a specified threshold, say $\gamma$, simultaneously.
Thus, we construct the following constrained optimization problem:
\begin{equation}
	\begin{aligned}\label{min:inject noise:original}
		\min_{\mSigma_k \in \sS} & \qquad \trace(\mSigma_k) \\
		\mathrm{s.t.} & \qquad \trace \big(\B(\vd_{k-1})\big) \geq \gamma, \ \mSigma_k \geq \sigma \mI_{\dx},
	\end{aligned}
\end{equation}
where $\B(\vd_{k - 1})$ represents the CRLB of $\vd_{k-1}$, $\gamma$ is a given value to quantify privacy level, and $\sigma$ is a small positive real number for numerical stability. The minimizer of \eqref{min:inject noise:original} is the sought-after $\mSigma_k$.  

We know from \eqref{eq:MSE of optimal estimator} that the greater $\mSigma_k$, the lower the state estimation accuracy. In addition, we know from \eqref{min:inject noise:original} that the greater $\gamma$, the higher the privacy level. 
Thus, by solving the problem \eqref{min:inject noise:original}, we  achieve the following two goals simultaneously:
\begin{itemize}
	\item \emph{Utility}: To minimize the MSE of the state estimate $\hat \vx_k$;
	\item \emph{Privacy}: To ensure that the privacy level of $\vd_{k - 1}$ is no less than a pre-set value $\gamma$.
\end{itemize}


Before solving the problem \eqref{min:inject noise:original}, we should first calculate $\B(\vd_{k - 1})$. It is worth noting that the state estimates employed by the adversary to estimate $\vd_{k - 1}$ are $\hat \vx_{k-\ns+1}, \hat \vx_{k-\ns+2}, \dots, \hat \vx_k$ with the latest state estimate $\hat \vx_k$ depending on $\mSigma_k$. Thus, $\B(\vd_{k - 1})$ is a function of $\mSigma_k$. We next provide an explicit expression for $\B(\vd_{k - 1})$ with respect to $\mSigma_k$.

\subsection{Explicit expression for CRLB}

It follows from \eqref{ineq:CRLB} that calculating the Fisher information matrix is the premise of calculating the CRLB.
Thus, we start by calculating the Fisher information matrix.

\textbf{\emph{1) Calculation for Fisher information matrix.}}
At time step $k$, we should calculate the Fisher information matrix for the history of exogenous inputs $\{\vd_0, \vd_1, \dots, \vd_{k - 1}\}$ based on the state estimates $\hat \vx_{k- \ns + 1}, \hat \vx_{k- \ns + 2}, \dots, \hat \vx_k$.  
Denote
\begin{align*}
	&\vd_{0 : k - 1} \coloneqq (\vd_0^{\Trans}, \vd_1^{\Trans}, \dots, \vd_{k - 1}^{\Trans})^{\Trans}, 
	\valpha_{k' : k} \coloneqq (\valpha_{k'}^{\Trans}, \valpha_{k'+1}^{\Trans}, \dots, \valpha_k^{\Trans})^{\Trans}, 
	\vx_{k' : k} \coloneqq (\vx_{k'}^{\Trans}, \vx_{k'+1}^{\Trans}, \dots, \hat \vx_k^{\Trans})^{\Trans}, \\
	&\hat \vx_{k' : k} \coloneqq (\hat \vx_{k'}^{\Trans}, \hat \vx_{k'+1}^{\Trans}, \dots, \hat \vx_k^{\Trans})^{\Trans}, 
	\hat \vx_{k' : k}^{\UMV} \coloneqq \Big(\big(\hat \vx_{k'}^{\UMV}\big)^{\Trans}, \big(\hat \vx_{k'+1}^{\UMV}\big)^{\Trans}, \dots, \big(\hat \vx_k^{\UMV}\big)^{\Trans}\Big)^{\Trans},
	k' \coloneqq k - \ns + 1.
\end{align*}
Then, we have
$\hat \vx_{k': k} = \hat \vx_{k' : k}^{\UMV} + \valpha_{k' : k}$.
Correspondingly, the covariance matrices of $\hat \vx_{k' : k} $ and $\hat \vx_{k' : k}^{\UMV}$, denoted by $\hat \mP_{k' : k}$ and $\hat \mP_{k' : k}^{\UMV}$, respectively, are correlated by
\begin{align}\label{eq:hat:P:aug}
	\hat \mP_{k' : k} &= \hat \mP_{k' : k}^{\UMV} + \mLambda_{\Sigma, k}, 
\end{align}
where $\mLambda_{\Sigma, k} = \blkdiag(\mSigma_{k'}, \mSigma_{k'+1}, \dots, \mSigma_k)$.
Due to the linearity of the system \eqref{eq:system} and the estimate \eqref{eq:UMV estimate}, we know that $\hat \vx_{k' : k}$ obeys a Gaussian distribution:
\begin{align}\label{distribution:data:Gaussian}
	\hat \vx_{k' : k} \sim \N(\E[\hat \vx_{k' : k}], \hat \mP_{k' : k}).
\end{align} 
Denote
\begin{align}\label{eq:Lk}
	&\mL_k = 
	\mL_{\DK, k}
	\mL_{\HF, k}
	\mLambda_{\G, k - 1},
\end{align}
where
\begin{align*}
	\mL_{\DK,k} &= 
	\begin{pmatrix}
		\prod_{i = 1}^{k'} \mD_i & \prod_{i = 2}^{k'} \mD_i & \dots & \mI_{\dx} \\
		\prod_{i = 1}^{k'+1} \mD_i & \prod_{i = 2}^{k'+1} \mD_i & \dots & \dots & \mI_{\dx} \\
		\vdots & \vdots & \vdots & \vdots & \vdots & \ddots \\
		\prod_{i = 1}^k \mD_i & \prod_{i = 2}^k \mD_i & \dots & \dots & \dots & \dots & \mI_{\dx}
	\end{pmatrix}  \mLambda_{\rK,k}, \\
	\mL_{\HF,k} &= \mLambda_{\rH,k} 
	 \begin{pmatrix}
		0 & 0 &\cdots & 0 & 0\\
		\mI_{\dx} & 0 &\cdots & 0 & 0 \\
		\prod_{i = 1}^1 \mF_i & \mI_{\dx} &\cdots & 0 & 0\\
		\vdots & \vdots & \ddots & \vdots &\vdots\\
		\prod_{i = 1}^{k-2} \mF_i & \prod_{i = 2}^{k-2} \mF_i&  \dots & \mI_{\dx} &  0 \\
		\prod_{i = 1}^{k - 1} \mF_i & \prod_{i = 2}^{k - 1} \mF_i & \dots & \mF_{k - 1} &  \mI_{\dx}
	\end{pmatrix}, \\
    \mD_k &= (\mI_{\dx} - \mK_k \mH_k) \mF_{k - 1}, \ \mLambda_{\rK,k} = \blkdiag(\mK_0, \mK_1, \dots, \mK_k), \\
	\mLambda_{\rH,k} &= \blkdiag(\mH_0, \mH_1, \dots, \mH_k),\ \mLambda_{\G,k-1} = \blkdiag(\mG_0, \mG_1, \dots, \mG_{k-1}).
\end{align*}
Then, we provide an explicit expression for the Fisher information matrix of $\vd_{0:k-1}$, as given in the following theorem.

\begin{theorem}\label{thm:Fisher information matrix}
	For the system \eqref{eq:system} and $X = \{\hat \vx_{k - \ns + 1}, \hat \vx_{k - \ns + 2}, \dots, \hat \vx_k\}$, 
	the Fisher information matrix of $\vd_{0:k-1}$ is given as
	\begin{align}\label{eq:EXPLICIT:FIM}
		\I(\vd_{0:k-1}) = \mL_k^{\Trans} \hat \mP_{k' : k}^{-1} \mL_k. 
	\end{align}
\end{theorem}

To provide a proof of Theorem \ref{thm:Fisher information matrix}, we need the following three lemmas, where the mean $\E[\hat \vx_{k' : k}]$ and the covariance matrix $\hat \mP_{k' : k}$ in \eqref{distribution:data:Gaussian} are expressed with respect to $\vd_{0:k - 1}$. They are the basis of the proof for Theorem \ref{thm:Fisher information matrix}.

\begin{lemma}[Fisher information matrix, \cite{Malago2015}]\label{lem:Fisher information matrix:Gaussian} 
	Let $\vx \sim \N(\vmu(\vtheta), \mSigma(\vtheta))$ be an $n$-variate Gaussian random vector with parameter $\vtheta = (\theta_1, \theta_2, \dots, \theta_m)^{\Trans}$. Then, for $1 \leq i, j \leq m$, the $(i,j)$ entry of the Fisher information matrix is
	\begin{align} \label{eq:FIM:Gaussian}
		\I_{i, j} = \frac{\partial \vmu}{\partial \theta_i} \mSigma^{-1} \frac{\partial \vmu^{\Trans}}{\partial \theta_j} + \frac{1}{2} \trace \left( \mSigma^{-1} \frac{\partial \mSigma}{\partial \theta_i} \mSigma^{-1} \frac{\partial \mSigma}{\partial \theta_j}\right).
	\end{align}
\end{lemma}

Based on Lemma \ref{lem:Fisher information matrix:Gaussian}, computing the Fisher information matrix  $\I(\vd_{0:k-1})$ requires first determining $\vmu = \E [\hat \vx_{k' : k}]$ and $\mSigma = \hat \mP_{k' : k}$, as presented in the following two lemmas.

\begin{lemma}\label{lem:Fisher:mean}
	$\E [\hat \vx_{k' : k}] = \mL_k \vd_{0:k-1} + \vc_k$ 
	with $\vc_k$ being a constant vector independent of $\vd_{0:k - 1}$.
\end{lemma}

\textit{Proof.}
Denote
$\vx_{0:k} \coloneqq
	(\vx_0^{\Trans},\vx_1^{\Trans},\dots,\vx_k^{\Trans})^{\Trans}, 
	\vy_{0:k} \coloneqq
(\vy_0^{\Trans},\vy_1^{\Trans},\dots,\vy_k^{\Trans})^{\Trans}, 
	\vw_{0:k - 1} \coloneqq
(\vw_0^{\Trans},\vw_1^{\Trans},\dots,\vw_{k - 1}^{\Trans})^{\Trans}$, and
$\vv_{0:k} \coloneqq
(\vv_0^{\Trans},\vv_1^{\Trans},\dots,\vv_k^{\Trans})^{\Trans}$.
Then, we have
\begin{align}
	\vx_{0:k} &= \mL_{\F, k - 1} \big(
		\vx_0^{\Trans},
		\vw_{0:k-1}^{\Trans}
	\big)^{\Trans} + \mL_{\tilde \F, k} \mLambda_{\G,k-1} \vd_{0:k-1}, \notag\\
	\vy_{0:k} 
	&= \mLambda_{\rH,k} \vx_{0:k} + \vv_{0:k}, \label{eq:y:aug}
\end{align}
where
\begin{align*}
	\mL_{\F,k - 1} &= 
	\begin{pmatrix}
		\mI_{\dx} &  & &  & \\
		\mF_0 & \mI_{\dx} & & &  \\
		\mF_1 \mF_0 &\mF_1 & \mI_{\dx} & & \\
		\vdots & \vdots & \vdots & \ddots & \\
		\prod_{i = 0}^{k - 1} \mF_i & \prod_{i = 1}^{k - 1} \mF_i & \prod_{i = 2}^{k - 1} \mF_i & \cdots &  \mI_{\dx}
	\end{pmatrix}, \ 
	\mL_{\tilde \F,k} &= 
	\begin{pmatrix}
		0 & 0 &\cdots & 0 & 0\\
		\mI_{\dx} & 0 &\cdots & 0 & 0 \\
		\prod_{i = 1}^1 \mF_i & \mI_{\dx} &\cdots & 0 & 0\\
		\vdots & \vdots & \ddots & \vdots &\vdots\\
		\prod_{i = 1}^{k-2} \mF_i & \prod_{i = 2}^{k-2} \mF_i&  \dots & \mI_{\dx} &  0 \\
		\prod_{i = 1}^{k - 1} \mF_i & \prod_{i = 2}^{k - 1} \mF_i & \dots & \mF_{k - 1} &  \mI_{\dx}
	\end{pmatrix}.
\end{align*}
Further, we have
\begin{align}
	\hat \vx_k^{\UMV} & = \mK_k \vy_k + \mD_k \hat \vx_{k-1}^{\UMV} \notag \\
	& = \mK_k \vy_k + \mD_k \mK_{k-1} \vy_{k-1} + \mD_k \mD_{k-1} \hat \vx_{k-2}^{\UMV} \notag \\
	&= \begin{pmatrix}
		\prod_{i = 1}^k \mD_i \mK_0, & \prod_{i = 2}^k \mD_i \mK_1, & \dots, & \mK_k
	\end{pmatrix}  \vy_{0:k}
	+ \prod_{i = 1}^k \mD_i (\mI_{\dx} - \mH_0 \mK_0) \bar \vx_0, \notag\\
	\hat \vx_{k' : k}^{\UMV} 
	&= \mL_{\DK, k}
	\vy_{0:k} + \begin{pmatrix}
		\big(\prod_{i = 1}^{k'} \mD_i\big)^{\Trans}, &
		\big(\prod_{i = 1}^{k'+1} \mD_i\big)^{\Trans},
		\dots, &
		\big(\prod_{i = 1}^k \mD_i\big)^{\Trans}
	\end{pmatrix}^{\Trans}
	(\mI_{\dx} - \mH_0 \mK_0) \bar \vx_0 \notag \\
	&= \mL_{\DK, k} 
	\mL_{\HF, k} \mLambda_{\G,k-1} \vd_{0:k-1}  + 
	\mL_{\DK, k} \Big(\mLambda_{\rH,k} \mL_{\F, k - 1} \big(
	\vx_0^{\Trans},
	\vw_{0:k-1}^{\Trans}
	\big)^{\Trans} + \vv_{0:k}\Big) \notag \\
	&\quad + \begin{pmatrix}
		\big(\prod_{i = 1}^{k'} \mD_i\big)^{\Trans}, &
		\big(\prod_{i = 1}^{k'+1} \mD_i\big)^{\Trans},
		\dots, &
		\big(\prod_{i = 1}^k \mD_i\big)^{\Trans}
	\end{pmatrix}^{\Trans}
	(\mI_{\dx} - \mH_0 \mK_0) \bar \vx_0, \notag \\
	&= \mL_{\DK, k} 
	\mL_{\HF, k} \mLambda_{\G,k-1} \vd_{0:k-1}  + 
	\mL_{\DK, k} \Big(\mLambda_{\rH,k} \mL_{\F, k - 1} \big(
	\vx_0^{\Trans},
	\vw_{0:k-1}^{\Trans}
	\big)^{\Trans} + \vv_{0:k}\Big) + \tilde \vc_k, \label{eq:hat:x:aug:loc}
\end{align}
where 
\begin{align*}
\tilde \vc_k = \begin{pmatrix}
	\big(\prod_{i = 1}^{k'} \mD_i\big)^{\Trans}, &
	\big(\prod_{i = 1}^{k'+1} \mD_i\big)^{\Trans},
	\dots, &
	\big(\prod_{i = 1}^k \mD_i\big)^{\Trans}
\end{pmatrix}^{\Trans}
(\mI_{\dx} - \mH_0 \mK_0) \bar \vx_0.
\end{align*}
Thus, we can obtain
$\E[\hat \vx_{k' : k}] = \E[\hat \vx_{k' : k}^{\UMV} + \valpha_{k' : k}] 
= \E[\hat \vx_{k' : k}^{\UMV}]
= \mL_k \vd_{0:k-1} + \vc_k$,
where 
\begin{align*}
\vc_k = \mL_{\DK, k} \Big(\mLambda_{\rH,k} \mL_{\F, k - 1} \big(
\bar \vx_0^{\Trans}, 0\big)^{\Trans} \Big) + \tilde \vc_k
\end{align*}
is a constant independent of $\vd_{0:k - 1}$. 
$\hfill\square$

\begin{lemma}\label{lem:Fisher:cov}
	$\hat \mP_{k' : k}$ is independent of $\vd_{0:k - 1}$. 
\end{lemma}

\textit{Proof.}
From \eqref{eq:y:aug} and the first equality of \eqref{eq:hat:x:aug:loc}, we have
\begin{align}\label{eq:hat:P:loc:aug}
	\hat \mP_{k' : k}^{\UMV} 
	&= \mL_{\DK, k} (\mLambda_{\rR, k} + \mLambda_{\rH, k} \mP_{0:k} \mLambda_{\rH, k}^{\Trans}) \mL_{\DK, k}^{\Trans}, 
\end{align}
where
\begin{align*}
\mP_{0:k} &= \E[(\vx_{0:k} - \E [\vx_{0:k}]) (\vx_{0:k} - \E [\vx_{0:k}])]^{\Trans}
= \mL_{\F, k - 1} \blkdiag(\mP_0, \mLambda_{\Q, k-1}) \mL_{\F, k - 1}^{\Trans}, \\
\mLambda_{\Q, k-1} &= \blkdiag(\mQ_0, \mQ_1, \dots, \mQ_{k-1}),
\mLambda_{\rR, k} = \blkdiag(\mR_0, \mR_1, \dots, \mR_k).
\end{align*}
We know from \eqref{eq:hat:P:loc:aug} that $\hat \mP_{k' : k}^{\UMV}$ is independent of $\vd_{k - 1}$. Moreover, from \eqref{eq:hat:P:aug} and the mutual independence between the perturbed noise and $\vd_{k - 1}$, we know that $\hat \mP_{k' : k}$ is also independent of $\vd_{0:k - 1}$. 
$\hfill\square$

Based on the above three lemmas, we now provide a proof of Theorem \ref{thm:Fisher information matrix}.

\textit{Proof of Theorem \ref{thm:Fisher information matrix}.}
	From \eqref{eq:FIM:Gaussian} and Lemmas \ref{lem:Fisher:mean}--\ref{lem:Fisher:cov}, by substituting $\vmu = \E [\hat \vx_{k' : k}]$ and $\mSigma = \hat \mP_{k' : k}$ into \eqref{eq:FIM:Gaussian}, 
	we have
	\begin{align*}
		\I(\vd_{0:k-1}) &= \frac{\partial \E [\hat \vx_{k' : k}]}{\partial \vd_{0:k - 1}} \hat \mP_{k' : k}^{-1} \frac{\partial \E [\hat \vx_{k' : k}^{\Trans}]}{\partial \vd_{0:k - 1}} 
		= \frac{\partial (\mL_k \vd_{0:k-1})}{\partial \vd_{0:k - 1}} \hat \mP_{k' : k}^{-1} \frac{\partial (\mL_k \vd_{0:k-1})^{\Trans}}{\partial \vd_{0:k - 1}} 
		= \mL_k^{\Trans} \hat \mP_{k' : k}^{-1} \mL_k,
	\end{align*}
	which completes the proof.
$\hfill\square$

Based on the explicit Fisher information matrix given by \eqref{eq:EXPLICIT:FIM}, we next provide an explicit expression for $\B(\vd_{k - 1})$.

\textbf{\emph{2) Calculation for CRLB.}}
At time step $k$, we focus on the privacy of $\vd_{k-1}$. Thus,  following the information inequality given by \eqref{ineq:CRLB}, we have
\begin{align}\label{eq:CRLB:information inequality}
	\B(\vd_{k-1}) 
	&= \bigg( \frac{\partial }{\partial \vd_{0:k-1}} g(\vd_{0:k-1}) \bigg)^{\Trans} \big( \mathcal I(\vd_{0:k-1}) \big)^{-1} 
	\frac{\partial }{\partial \vd_{0:k-1}} g(\vd_{0:k-1})  \notag \\
	&= \begin{pmatrix}
		0& \mI_{\dd}	
	\end{pmatrix} \big(\mL_k^{\Trans} \hat \mP_{k' : k}^{-1} \mL_k\big)^{-1} \begin{pmatrix}
		0& \mI_{\dd}	
	\end{pmatrix}^{\Trans},
\end{align}
where $g(\vd_{0:k-1}) = \vd_{k-1}$ with $g$ being a projection operator.
We know from \eqref{eq:CRLB:information inequality} that $\B(\vd_{k-1})$ is the $\dd \times \dd$ block at the right bottom of the inverse of the Fisher information matrix given by \eqref{eq:EXPLICIT:FIM}. 
However, \eqref{eq:CRLB:information inequality} cannot be employed directly since it is implicit with respect to the optimization variable $\mSigma_k$. In fact, adopting \eqref{eq:CRLB:information inequality} directly will make \eqref{min:inject noise:original} difficult to solve.
Therefore, we need to simplify \eqref{eq:CRLB:information inequality} further to an explicit form with respect to $\mSigma_k$. 
By denoting the following matrices as partitioned forms:
\begin{align}\label{eq:hat P:UMV:aug}
	\hat \mP_{k' : k}^{\UMV} =  \begin{pmatrix}
		\hat \mP_{k' : k - 1}^{\UMV} & \hat \mP_{k' : k - 1, k}^{\UMV}\\
		\hat \mP_{k, k' : k - 1}^{\UMV} & \hat \mP_k^{\UMV}
	\end{pmatrix}, 
	\hat \mP_{k' : k} = \begin{pmatrix}
		\hat \mP_{k' : k - 1} & \hat \mP_{k' : k - 1, k}^{\UMV}\\
		\hat \mP_{k, k' : k - 1}^{\UMV} & \hat \mP_k^{\UMV} + \mSigma_k
	\end{pmatrix}, 
	\mL_k = \begin{pmatrix}
		\mL_{11}(k) & 0\\
		\mL_{21}(k) & \mG_{k-1}
	\end{pmatrix}, 
\end{align}
where $\mL_{11}(k) \in \R^{(\ns - 1) \dx \times (k - 1) \dd}$, $\mL_{21}(k) \in \R^{\dx \times (k - 1) \dd}$,
we then provide an explicit expression for $\B(\vd_{k - 1})$ with respect to $\mSigma_k$ in the following theorem.

\begin{theorem}\label{lem:CRLB computation}
	For the system \eqref{eq:system} and $X = \{\hat \vx_{k - \ns + 1}, \hat \vx_{k - \ns + 2}, \dots, \hat \vx_k\}$, an explicit expression for $\B(\vd_{k - 1})$ with respect to $\mSigma_k$ is given by
	\begin{align}\label{eq:explicit CRLB}
		\B(\vd_{k - 1}) &= \big( \mG_{k-1}^{\Trans} (\mSigma_k + \mA_k)^{-1} \mG_{k-1}\big)^{-1},
	\end{align}
	where 
	\begin{align}
		\mA_k &=  \hat \mP_k^{\UMV} - \hat \mP_{k, k':k - 1}^{\UMV} \hat \mP_{k':k - 1}^{-1} 
		\hat \mP_{k':k - 1, k}^{\UMV}  + \big(\mL_{21}(k) -  \hat \mP_{k, k':k - 1}^{\UMV} 
		\hat \mP_{k':k - 1}^{-1} \mL_{11}(k)\big)  \notag\\
		&\quad \cdot
		\big(\mL_{11}(k)^{\Trans}  \hat \mP_{k':k - 1}^{-1} \mL_{11}(k)\big)^{-1} \big(\mL_{21}(k)^{\Trans} - \mL_{11}(k)^{\Trans} \hat \mP_{k':k - 1}^{-1} \hat \mP_{k':k - 1, k}^{\UMV}\big).\label{eq:Ak}
	\end{align}
\end{theorem}

\textit{Proof.}
	 Denote
$
\hat \mP_{k' : k}^{-1}
= \begin{pmatrix}
	\mGamma_{11} & \mGamma_{12} \\
	\mGamma_{21} & \mGamma_{22}
\end{pmatrix}.
$
	Then, we have
	\begin{align*} 
		\mGamma_{11} &= \hat \mP_{k' : k - 1}^{-1} + \hat \mP_{k' : k - 1}^{-1} \hat \mP_{k' : k - 1, k}^{\UMV} \mDelta^{-1} 
		\hat \mP_{k, k' : k - 1}^{\UMV} \hat \mP_{k' : k - 1}^{-1}, 
		\mDelta = \hat \mP_k^{\UMV} + \mSigma_k - \hat \mP_{k, k' : k - 1}^{\UMV} \hat \mP_{k' : k - 1}^{-1} 
		\hat \mP_{k' : k - 1, k}^{\UMV}, \\
		\mGamma_{12} &=  -\hat \mP_{k' : k - 1}^{-1} \hat \mP_{k' : k - 1, k}^{\UMV} \mDelta^{-1}, \ 
		\mGamma_{21} = -\mDelta^{-1} \hat \mP_{k, k' : k - 1}^{\UMV} \hat \mP_{k' : k - 1}^{-1}, \ 
		\mGamma_{22} = \mDelta^{-1}. 		
	\end{align*}
	Further, denote
	\begin{align*}
		\mL_k^{\Trans} \hat \mP_{k' : k}^{-1} \mL_k  
		&=
		\begin{pmatrix}
			\mL_{11}(k)^{\Trans} & \mL_{21}(k)^{\Trans}\\
			0                 & \mG_{k-1}^{\Trans} 
		\end{pmatrix}
		\begin{pmatrix}
			\mGamma_{11} & \mGamma_{12} \\
			\mGamma_{21} & \mGamma_{22}
		\end{pmatrix}
		\begin{pmatrix}
			\mL_{11}(k) & 0 \\
			\mL_{21}(k) & \mG_{k-1}
		\end{pmatrix} 
		= \begin{pmatrix}
			\mPhi_{11} & \mPhi_{12}\\
			\mPhi_{21} & \mPhi_{22}
		\end{pmatrix}.
	\end{align*}
	Then, we have
	\begin{align*}
		\mPhi_{11} 
		&= \mL_{11}(k)^{\Trans} \hat \mP_{k' : k - 1}^{-1} \mL_{11}(k) 
		+ \mL_{11}(k)^{\Trans} \hat \mP_{k' : k - 1}^{-1}  \hat \mP_{k' : k - 1, k}^{\UMV} \mDelta^{-1} \hat \mP_{k, k' : k - 1}^{\UMV} \hat \mP_{k' : k - 1}^{-1} 
		\mL_{11}(k)
		- \mL_{21}(k)^{\Trans} \mDelta^{-1} \\
		&\quad \cdot \hat \mP_{k, k':k - 1}^{\UMV} \hat \mP_{k' : k - 1}^{-1} \mL_{11}(k)  + \mL_{21}(k)^{\Trans} \mDelta^{-1} \mL_{21}(k) 
		- \mL_{11}(k)^{\Trans} \hat \mP_{k':k - 1} ^{-1} 
		\hat \mP_{k':k - 1, k}^{\UMV} \mDelta^{-1} \mL_{21}(k), \\
		\mPhi_{12} &= \mL_{21}(k)^{\Trans} \mDelta^{-1} \mG_{k-1} - \mL_{11}(k)^{\Trans} \hat \mP_{k':k - 1}^{-1} \hat \mP_{k':k - 1, k}^{\UMV} 
		\mDelta^{-1} \mG_{k-1}, \\
		\mPhi_{21} &= \mG_{k-1}^{\Trans} \mDelta^{-1} \mL_{21}(k) - \mG_{k-1}^{\Trans} \mDelta^{-1} \hat \mP_{k, k':k - 1}^{\UMV} 
		\hat \mP_{k':k - 1}^{-1} \mL_{11}(k), 
		\mPhi_{22} = \mG_{k-1}^{\Trans} \mDelta^{-1} \mG_{k-1}.
	\end{align*}
	Substituting into \eqref{eq:CRLB:information inequality}, we have
	\begin{align*}
		\B(\vd_{k - 1}) 
		&= \begin{pmatrix}
			0& \mI_{\dd}	
		\end{pmatrix} \begin{pmatrix}
		\mPhi_{11} & \mPhi_{12}\\
		\mPhi_{21} & \mPhi_{22}
	\end{pmatrix}^{-1} \begin{pmatrix}
			0& \mI_{\dd}	
		\end{pmatrix}^{\Trans} \notag \\
		&= \Big(\mG_{k-1}^{\Trans} \mDelta^{-1} \mG_{k-1}- \big(\mG_{k-1}^{\Trans} \mDelta^{-1} \mL_{21}(k)  - \mG_{k-1}^{\Trans} \mDelta^{-1} 
		\hat \mP_{k, k':k - 1}^{\UMV} \hat \mP_{k':k - 1}^{-1} \mL_{11}(k)\big) \notag\\
		&\quad \cdot \big(\mL_{11}(k)^{\Trans} \hat \mP_{k':k - 1}^{-1} \mL_{11}(k)  + \mL_{11}(k)^{\Trans} \hat \mP_{k':k - 1}^{-1} 
		\hat \mP_{k':k - 1, k}^{\UMV} \mDelta^{-1} \hat \mP_{k, k':k - 1}^{\UMV} \hat \mP_{k':k - 1}^{-1} \mL_{11}(k) \notag\\
		&\quad - \mL_{21}(k)^{\Trans} \mDelta^{-1} \hat \mP_{k, k':k - 1}^{\UMV} \hat \mP_{k':k - 1}^{-1} \mL_{11}(k)  + \mL_{21}(k)^{\Trans} 
		\mDelta^{-1} \mL_{21}(k)  - \mL_{11}(k)^{\Trans} \hat \mP_{k':k - 1}^{-1} \notag\\
		&\quad \cdot
		\hat \mP_{k':k - 1, k}^{\UMV} \mDelta^{-1} 
		\mL_{21}(k)\big)^{-1} 
		\big(\mL_{21}(k)^{\Trans} \mDelta^{-1} \mG_{k-1}- \mL_{11}(k)^{\Trans} \hat \mP_{k':k - 1}^{-1} 
		\hat \mP_{k':k - 1, k}^{\UMV} \mDelta^{-1} \mG_{k-1}\big)\Big)^{-1}.
	\end{align*}
	By some simplifications, we then have
	\begin{align}\label{eq:tmp:1}
		\B(\vd_{k - 1}) 
		&= \Bigg( \mG_{k-1}^{\Trans} \bigg( \mDelta^{-1} - \mDelta^{-1} \Big( \mL_{21}(k) -  \hat \mP_{k, k':k - 1}^{\UMV} 
		\hat \mP_{k':k - 1}^{-1} \mL_{11}(k) \Big)  \Big( \mL_{11}(k)^{\Trans}  \hat \mP_{k':k - 1}^{-1} \mL_{11}(k) \notag\\
		&\quad + \big(\mL_{21}(k)^{\Trans} - \mL_{11}(k)^{\Trans} \hat \mP_{k':k - 1}^{-1} \hat \mP_{k, k':k - 1}^{\UMV}\big) \mDelta^{-1} \big(\mL_{21}(k) - \hat \mP_{k, k':k - 1}^{\UMV} \hat \mP_{k':k - 1}^{-1} \mL_{11}(k)\big) \Big)^{-1} \notag\\
		&\quad \cdot \Big(\mL_{21}(k)^{\Trans} - \mL_{11}(k)^{\Trans} \hat \mP_{k':k - 1}^{-1} \hat \mP_{k':k - 1, k}^{
			\UMV}\Big) \mDelta^{-1} \bigg) 
		\mG_{k-1}\Bigg)^{-1}. 
	\end{align}
	Using the Woodbury matrix identity (see Page 258 of \cite{Higham2002Accuracy}) for the right-hand side of \eqref{eq:tmp:1}, we have
	\begin{align*}
		\B(\vd_{k - 1}) 
		& = \bigg( \mG_{k-1}^{\Trans} \Big( \mDelta + \big(\mL_{21}(k) -  \hat \mP_{k, k':k - 1}^{\UMV} \hat \mP_{k':k - 1}^{-1} 
		\mL_{11}(k)\big)  \big(\mL_{11}(k)^{\Trans}  \hat \mP_{k':k - 1}^{-1} \mL_{11}(k)\big)^{-1} \notag \\
		&\quad \cdot \big(\mL_{21}(k)^{\Trans} - \mL_{11}(k)^{\Trans} 
		\hat \mP_{k':k - 1}^{-1} \hat \mP_{k':k - 1, k}^{\UMV}\big) \Big)^{-1} 
		\mG_{k-1}\bigg)^{-1} \notag\\
		&= \bigg( \mG_{k-1}^{\Trans} \Big( \hat \mP_k^{\UMV} + \mSigma_k - \hat \mP_{k, k':k - 1}^{\UMV} \hat \mP_{k':k - 1}^{-1} 
		\hat \mP_{ k':k - 1, k}^{\UMV}  + \big(\mL_{21}(k) -  \hat \mP_{k,  k':k - 1}^{\UMV} 
		\hat \mP_{k':k - 1}^{-1} \mL_{11}(k)\big) \\
		&\quad \cdot \big(\mL_{11}(k)^{\Trans}  \hat \mP_{k':k - 1}^{-1} \mL_{11}(k)\big)^{-1}   \big(\mL_{21}(k)^{\Trans} - \mL_{11}(k)^{\Trans} \hat \mP_{k':k - 1}^{-1} \hat \mP_{k':k - 1, k}^{\UMV}\big) \Big)^{-1} \mG_{k-1}\bigg)^{-1} \\
		&= \big( \mG_{k-1}^{\Trans} (\mSigma_k + \mA_k)^{-1} \mG_{k-1}\big)^{-1},
	\end{align*}
where $\mA_k$ is given by \eqref{eq:Ak}. 
$\hfill\square$	

Based on the explicit expression for $\B(\vd_{k - 1})$ given in Theorem \ref{lem:CRLB computation}, the optimization problem \eqref{min:inject noise:original} can be rewritten more clearly as follows:
\begin{equation}
	\begin{aligned}\label{min:inject noise:0}
		\min_{\mSigma_k \in \sS} & \qquad \trace(\mSigma_k) \\
		\mathrm{s.t.} & \qquad \trace \Big(\big( \mG_{k-1}^{\Trans} (\mSigma_k + \mA_k)^{-1} \mG_{k-1}\big)^{-1}\Big) \geq \gamma, \ \mSigma_k \geq \sigma \mI_{\dx}.
	\end{aligned}
\end{equation}

\begin{remark}
	As suggested by \eqref{min:inject noise:0}, a larger value of $\gamma$ tends to result in a larger matrix $\mSigma_k$ in the L\"owner order sense. 
	The magnitude of $\gamma$ reflects the strength of privacy protection. A larger  $\gamma$ corresponds to stronger privacy, which requires increasing the uncertainty of the added noise and consequently reduces the accuracy of state estimation.  
	In another word, enhanced privacy protection is often achieved at the expense of state estimation accuracy.
	This trade-off is a common characteristic of privacy-preserving methods based on random perturbations. For example, the same principle applies to differential privacy: smaller $\epsilon$ and $\delta$ values correspond to stronger privacy and require greater noise uncertainty, ultimately at the expense of estimation accuracy (see, e.g., \cite{LeNy2014}). 
	The specific choice of  $\gamma$ depends on the requirements of the practical problem.
\end{remark}

However, there exist two substantial difficulties in solving \eqref{min:inject noise:0}. 
Firstly, calculating $\B(\vd_{k - 1})$ directly suffers from a heavy computational burden since the computational complexity of $\B(\vd_{k - 1})$ becomes greater over the time step $k$.
More specifically, the computational complexity of the matrix $\mA_k$ given by \eqref{eq:Ak} increases over the time step $k$.
Essentially, this is caused by the fact that calculating $\B(\vd_{k - 1})$ requires the history of all the measurements up to time step $k$, i.e., $\vy_0, \vy_1, \dots, \vy_k$. 
As such, the computational complexity of $\B(\vd_{k - 1})$ tends to become more expensive over time step $k$.
Secondly, \eqref{min:inject noise:0} is a non-convex optimization problem since its first constraint is non-convex. Thus, an analytic solution of \eqref{min:inject noise:0} is hard to obtain. 
As well known, computational efficiency is critically important for real-time state estimation. Thus, reducing the computational complexity of $\B(\vd_{k - 1})$ and solving \eqref{min:inject noise:0} efficiently are the two important issues that we will address sequentially in the following section.

\section{Privacy-preserving state estimation with low complexity}\label{sec:online state estimation}

In this section, we first reduce the computational complexity of $\B(\vd_{k - 1})$. Then, we propose a relaxation approach for solving \eqref{min:inject noise:0} and provide the privacy-preserving state estimation algorithm with low complexity. Finally, we show that the proposed algorithm also ensures $(\epsilon, \delta)$-differential privacy.

\subsection{Privacy-preserving state estimation algorithm with low complexity}

For calculating the $\B(\vd_{k - 1})$ given by \eqref{eq:explicit CRLB}, the main computation lies in $\mA_k$  given by \eqref{eq:Ak}.
As the computational complexity of \eqref{eq:Ak} mainly lies in the matrices $\hat \mP_{k' : k}^{\UMV}$ given by \eqref{eq:hat:P:loc:aug} and $\mL_k$ given by \eqref{eq:Lk}, we next provide low-complexity calculations for these matrices.

\textbf{\emph{1) Recursive calculation for $\hat \mP_{k' : k}^{\UMV}$.}} 
To avoid directly calculating the matrix $\hat \mP_{k' : k}^{\UMV}$ at each time step $k$, we design a recursive calculation for $\hat \mP_{k' : k}^{\UMV}$ as follows.

\begin{theorem}\label{thm:recursive computation of P}
	The following recursions hold:
	\begin{align}
		\Cov(\vx_0, \vx_0) &= \mP_0, \notag \\
		\Cov(\vx_0, \hat \vx_0^{\UMV}) &= \mP_0 \mH_0^{\Trans} \mK_0^{\Trans}, \label{eq:cov:x0 and hat x0}\\
		\Cov(\hat \vx_0^{\UMV}, \hat \vx_0^{\UMV}) &= \mK_0 \mH_0 \mP_0 \mH_0^{\Trans} \mK_0^{\Trans} + \mK_0 \mR_0 \mK_0^{\Trans}.\label{eq:cov:hat x0 and hat x0}
	\end{align}
	For $k = 1, 2, \dots$,
	\begin{align}
		\Cov(\vx_k, \vx_k) 
		&= \mF_{k - 1} \Cov(\vx_{k - 1}, \vx_{k - 1}) \mF_{k - 1}^{\Trans} 
		+ \mQ_{k - 1}, \label{eq:cov:xk and xk}\\
		\Cov(\hat \vx_k^{\UMV}, \hat \vx_k^{\UMV}) &= \mD_k \Cov(\hat \vx_{k - 1}^{\UMV}, \hat \vx_{k - 1}^{\UMV}) \mD_k^{\Trans}  + \mD_k \Cov(\hat \vx_{k - 1}^{\UMV}, \vx_{k - 1}) \mF_{k - 1}^{\Trans} \mH_k^{\Trans} \mK_k^{\Trans} \notag \\
		&\quad + \mK_k \mH_k \mF_{k - 1} \Cov(\vx_{k - 1}, \hat \vx_{k - 1}^{\UMV}) \mD_k^{\Trans}  + \mK_k \mH_k \Cov(\vx_k, \vx_k) \mH_k^{\Trans} \mK_k^{\Trans}  + \mK_k \mR_k \mK_k^{\Trans}, \label{eq:cov:xk hat and xk hat}\\	
		\Cov(\vx_k, \hat \vx_k^{\UMV}) 
		&= \mF_{k - 1} \Cov(\vx_{k - 1}, \hat \vx_{k - 1}^{\UMV}) \mD_k^{\Trans}  + \Cov(\vx_k, \vx_k) \mH_k^{\Trans} \mK_k^{\Trans}. \label{eq:cov:xk hat and xk}
	\end{align}
	For $j \in \Z$, we have
	\begin{align}
		\Cov(\vx_{k + j}, \hat \vx_k^{\UMV}) 
		&= \mF_{k + j - 1} \Cov(\vx_{k + j - 1}, \hat \vx_k^{\UMV}), \label{eq:cov:xk hat and xk+j} \\
		\Cov(\hat \vx_{k + j}^{\UMV}, \hat \vx_k^{\UMV}) 
		&= \mD_{k + j} \Cov(\hat \vx_{k + j - 1}^{\UMV}, \hat \vx_k^{\UMV}) 
		+ \mK_{k + j} \mH_{k + j} \mF_{k + j - 1} \Cov(\vx_{k + j - 1}, \hat \vx_k^{\UMV}). \label{eq:cov:hat xk and xk+j hat}
	\end{align}
\end{theorem}

\textit{Proof.}
	To complete this proof, we derive each of \eqref{eq:cov:x0 and hat x0}--\eqref{eq:cov:hat xk and xk+j hat} as follows. 
	For \eqref{eq:cov:x0 and hat x0}, we have
	\begin{align*}
		\Cov(\vx_0, \hat \vx_0^{\UMV}) 
		= \Cov\big(\vx_0, \bar \vx_0 + \mK_0 (\vy_0 - \mH_0 \bar \vx_0)\big) 
		= \Cov(\vx_0, \mK_0 \vy_0) 
		= \Cov(\vx_0, \mK_0 \mH_0 \vx_0) 
		= \mP_0 \mH_0^{\Trans} \mK_0^{\Trans}.
	\end{align*}
	For \eqref{eq:cov:hat x0 and hat x0}, we have
	\begin{align*}
		\Cov(\hat \vx_0^{\UMV}, \hat \vx_0^{\UMV}) 
		&= \Cov\big(\bar \vx_0 + \mK_0 (\vy_0 - \mH_0 \bar \vx_0), \bar \vx_0 + \mK_0 (\vy_0 - \mH_0 \bar \vx_0)\big) \\
		&= \Cov\big(\mK_0 (\vy_0 - \mH_0 \bar \vx_0), \mK_0 (\vy_0 - \mH_0 \bar \vx_0) \big) \\
		&= \Cov(\mK_0 \vy_0, \mK_0 \vy_0) \\
		&= \Cov\big(\mK_0 (\mH_0 \vx_0 + \vv_0), \mK_0 (\mH_0 \vx_0 + \vv_0)\big) \\
		&= \mK_0 \mH_0 \mP_0 \mH_0^{\Trans} \mK_0^{\Trans} + \mK_0 \mR_0 \mK_0^{\Trans}.
	\end{align*}
	For \eqref{eq:cov:xk and xk}, we have
	\begin{align*}
		\Cov(\vx_k, \vx_k) 
		&= \Cov(\mF_{k - 1} \vx_{k - 1} + \mG_{k - 1} \vd_{k - 1} + \vw_{k - 1},  \mF_{k - 1} \vx_{k - 1} + \mG_{k - 1} \vd_{k - 1} + \vw_{k - 1}) \\
		&= \mF_{k - 1} \Cov(\vx_{k - 1}, \vx_{k - 1}) \mF_{k - 1}^{\Trans} + \mQ_{k - 1}.
	\end{align*} 
	For \eqref{eq:cov:xk hat and xk hat}, we have
	\begin{align*}
		&\Cov(\hat \vx_k^{\UMV}, \hat \vx_k^{\UMV}) \\
		&= \Cov(\mD_k \hat \vx_{k - 1}^{\UMV} + \mK_k \vy_k, \mD_k \hat \vx_{k - 1}^{\UMV} + \mK_k \vy_k) \\
		&= \Cov(\mD_k \hat \vx_{k - 1}^{\UMV}, \mD_k \hat \vx_{k - 1}^{\UMV}) + \Cov(\mD_k \hat \vx_{k - 1}^{\UMV}, \mK_k \vy_k)  + \Cov(\mK_k \vy_k, \mD_k \hat \vx_{k - 1}^{\UMV}) + \Cov(\mK_k \vy_k, \mK_k \vy_k) \\
		&= \mD_k \Cov(\hat \vx_{k - 1}^{\UMV}, \hat \vx_{k - 1}^{\UMV}) \mD_k^{\Trans}  + \Cov\big(\mD_k \hat \vx_{k - 1}^{\UMV}, \mK_k(\mH_k \vx_k + \vv_k)\big)  + \Cov\big(\mK_k(\mH_k \vx_k + \vv_k), \mD_k \hat \vx_{k - 1}^{\UMV}\big) \\
		&\quad + \Cov\big(\mK_k(\mH_k \vx_k + \vv_k), \mK_k(\mH_k \vx_k + \vv_k)\big) \\
		&= \mD_k \Cov(\hat \vx_{k - 1}^{\UMV}, \hat \vx_{k - 1}^{\UMV}) \mD_k^{\Trans}  + \Cov(\mD_k \hat \vx_{k - 1}^{\UMV}, \mK_k \mH_k \vx_k)  + \Cov(\mK_k \mH_k \vx_k, \mD_k \hat \vx_{k - 1}^{\UMV}) \\
		&\quad + \Cov(\mK_k \mH_k \vx_k, \mK_k \mH_k \vx_k)  + \Cov(\mK_k \vv_k, \mK_k \vv_k) \\
		&= \mD_k \Cov(\hat \vx_{k - 1}^{\UMV}, \hat \vx_{k - 1}^{\UMV}) \mD_k^{\Trans}  + \mD_k \Cov(\hat \vx_{k - 1}^{\UMV}, \vx_{k - 1}) \mF_{k - 1}^{\Trans} \mH_k^{\Trans} \mK_k^{\Trans}  + \mK_k \mH_k \mF_{k - 1} \Cov(\vx_{k - 1}, \hat \vx_{k - 1}^{\UMV}) \mD_k^{\Trans} \\
		&\quad + \mK_k \mH_k \Cov(\vx_k, \vx_k) \mH_k^{\Trans} \mK_k^{\Trans} + \mK_k \mR_k \mK_k^{\Trans}.
	\end{align*}
	For \eqref{eq:cov:xk hat and xk}, we have
	\begin{align*}
		\Cov(\vx_k, \hat \vx_k^{\UMV}) 
		&= \Cov(\vx_k, \mD_k \hat \vx_{k - 1}^{\UMV} + \mK_k \vy_k) \\
		&= \Cov(\vx_k, \hat \vx_{k - 1}^{\UMV}) \mD_k^{\Trans} + \Cov(\vx_k, \vy_k) \mK_k^{\Trans} \\
		&= \Cov(\vx_k, \hat \vx_{k - 1}^{\UMV}) \mD_k^{\Trans} +\Cov(\vx_k, \vx_k) \mH_k^{\Trans} \mK_k^{\Trans} \\
		&= \mF_{k - 1} \Cov(\vx_{k - 1}, \hat \vx_{k - 1}^{\UMV}) \mD_k^{\Trans} + \Cov(\vx_k, \vx_k) \mH_k^{\Trans} \mK_k^{\Trans}.
	\end{align*}
	For \eqref{eq:cov:xk hat and xk+j}, we have
	\begin{align*}
		\Cov(\vx_{k + j}, \hat \vx_k^{\UMV}) 
		&= \Cov(\mF_{k + j - 1} \vx_{k + j - 1} + \mG_{k + j -1} \vd_{k + j - 1}  + \vw_{k + j - 1}, \hat \vx_k^{\UMV}) \\
		&= \mF_{k + j - 1} \Cov(\vx_{k + j - 1}, \hat \vx_k^{\UMV}).
	\end{align*}
	For \eqref{eq:cov:hat xk and xk+j hat}, we have
	\begin{align*}
		\Cov(\hat \vx_{k + j}^{\UMV}, \hat \vx_k^{\UMV}) 
		&= \Cov(\mD_{k + j} \hat \vx_{k + j - 1}^{\UMV} + \mK_{k + j} \vy_{k + j}, \hat \vx_k^{\UMV}) \\
		&= \mD_{k + j} \Cov(\hat \vx_{k + j - 1}^{\UMV}, \hat \vx_k^{\UMV})  + \mK_{k + j} \Cov(\vy_{k + j}, \hat \vx_k^{\UMV}) \\
		&= \mD_{k + j} \Cov(\hat \vx_{k + j - 1}^{\UMV}, \hat \vx_k^{\UMV})  + \mK_{k + j} \Cov(\mH_{k + j} \vx_{k + j} + \mH_{k + j} \vv_{k + j}, \hat \vx_k^{\UMV}) \\
		&= \mD_{k + j} \Cov(\hat \vx_{k + j - 1}^{\UMV}, \hat \vx_k^{\UMV})  + \mK_{k + j} \mH_{k + j} \Cov(\vx_{k + j}, \hat \vx_k^{\UMV}) \\
		&= \mD_{k + j} \Cov(\hat \vx_{k + j - 1}^{\UMV}, \hat \vx_k^{\UMV})  + \mK_{k + j} \mH_{k + j} \mF_{k + j - 1} \Cov(\vx_{k + j - 1}, \hat \vx_k^{\UMV}).
	\end{align*}
The proof is completed.
$\hfill\square$

Following Theorem \ref{thm:recursive computation of P}, we realize recursive calculation for $\hat \mP_{k' : k}^{\UMV}$ by computing each sub-block of $\hat \mP_{k' : k}^{\UMV}$ recursively at each time step $k$.

\textbf{\emph{2) Time-independent calculation for $\mL_k$.}}
We can see from \eqref{eq:hat P:UMV:aug} that the computational complexity of $\mL_k$ is caused by $\mL_{11}(k)$ and $\mL_{21}(k)$ since the sizes of these two matrices are dependent on the time step $k$. To solve this problem, we introduce the following pseudo-CRLB (PCRLB) by replacing $\mL_k$ with $\tilde \mL_k$ in \eqref{eq:CRLB:information inequality}:
\begin{align}\label{eq:PCRLB}
	\PB(\vd_{k - 1}) &\coloneqq 
	\begin{pmatrix}
		0& \mI_{\dd}	
	\end{pmatrix} \Big(\tilde \mL_k^{\Trans} \hat \mP_{k' : k}^{-1} \tilde \mL_k\Big)^{-1} \begin{pmatrix}
		0& \mI_{\dd}	
	\end{pmatrix}^{\Trans},
\end{align}
where
$\tilde \mL_k = \mL_k \begin{pmatrix}
	0 & \mI_{\ns \dd}
\end{pmatrix}^{\Trans} \in \R^{\ns \dx \times \ns \dd}$. 
It is worth noting that the size of $\tilde \mL_k$ is independent of the time step $k$.
To calculate $\tilde \mL_k$ (instead of calculating $\mL_k$), denote
\begin{align}
	\tilde \mL_{\DK, k} 
	&= \begin{pmatrix}
		\mI_{\dx} \\
		\mD_{k'+1} & \mI_{\dx} \\
		\prod_{i = k'+1}^{k'+2} \mD_i & \mD_{k'+2} & \mI_{\dx} \\
		\vdots & \vdots & \vdots & \ddots \\
		\prod_{i = k'+1}^k \mD_i & \prod_{i = k'+2}^k \mD_i & \dots & \dots & \mI_{\dx}
	\end{pmatrix}  \blkdiag(\mK_{k'}, \mK_{k'+1}, \dots, \mK_k), \label{eq:tilde LDK}\\
	\tilde \mL_{\HF, k} &= \blkdiag(\mH_{k'}, \mH_{k'+1}, \dots, \mH_k) 
	\begin{pmatrix}
		\mI_{\dx} \\
		\mF_{k'} & \mI_{\dx} \\
		\prod_{i = k'}^{k'+1} \mF_i & \mF_{k'+1} & \mI_{\dx} \\
		\prod_{i = k'}^{k'+2} \mF_i & \prod_{i = k'+1}^{k'+2} \mF_i & \mF_{k'+2} & \mI_{\dx} \\
		\vdots & \vdots & \vdots & \vdots & \ddots \\
		\prod_{i = k'}^{k - 1} \mF_i & \prod_{i = k'+1}^{k - 1} \mF_i & \dots & \dots & \dots & \mI_{\dx}
	\end{pmatrix}. \label{eq:tilde LHF}
\end{align}
Then, the following proposition presents the calculation for $\tilde \mL_k$.

\begin{proposition}\label{prop:tilde:Lk:computation}
	$\tilde \mL_k$ can be calculated as follows:
	\begin{align}\label{eq:explicit:tilde Lk}
		\tilde \mL_k &= \tilde \mL_{\DK, k} \tilde \mL_{\HF, k} \blkdiag(\mG_{k - \ns}, \dots, \mG_{k - 1}).
	\end{align}
\end{proposition}

\textit{Proof.}
	From \eqref{eq:Lk} and $\tilde \mL_k = \mL_k \begin{pmatrix}
		0 & \mI_{\ns \dd}
	\end{pmatrix}^{\Trans}$, we have
	\begin{align*}
		\tilde \mL_k &= \mL_{\DK, k}
		\mL_{\HF, k}
		\mLambda_{\G, k - 1} \begin{pmatrix}
			0 & \mI_{\ns \dd}
		\end{pmatrix}^{\Trans} \notag \\
		&= \mL_{\DK, k}
		\mL_{\HF, k} \begin{pmatrix}
			0 \\
			\blkdiag(\mG_{k - \ns}, \dots, \mG_{k - 1}) 
		\end{pmatrix} \notag\\
		&= \mL_{\DK, k} \begin{pmatrix}
			0 \\
			\tilde \mL_{\HF, k} \blkdiag(\mG_{k - \ns}, \dots, \mG_{k - 1}) 
		\end{pmatrix} \notag\\
		&= \tilde \mL_{\DK, k} \tilde \mL_{\HF, k} \blkdiag(\mG_{k - \ns}, \dots, \mG_{k - 1}).
	\end{align*}
	This completes the proof.
$\hfill\square$

The following proposition provides an explicit expression for the $\PB(\vd_{k - 1})$.

\begin{proposition}\label{prop:explicit:PCRLB}
	An explicit expression for the $\PB(\vd_{k - 1})$ with respect to $\mSigma_k$ is given by
	\begin{align}\label{eq:explicit PCRLB}
		\PB(\vd_{k - 1}) = \big( \mG_{k-1}^{\Trans} (\mSigma_k + \tilde \mA_k)^{-1} \mG_{k-1}\big)^{-1},
	\end{align}
	where 
	\begin{align}
		\tilde \mA_k &=  \hat \mP_k^{\UMV} - \hat \mP_{k, k':k - 1}^{\UMV} \hat \mP_{k':k - 1}^{-1} 
		\hat \mP_{k':k - 1, k}^{\UMV} + \big(\tilde \mL_{21} -  \hat \mP_{k, k':k - 1}^{\UMV} 
		\hat \mP_{k':k - 1}^{-1} \tilde \mL_{11}\big) \notag\\
		&\quad \cdot 
		\big(\tilde \mL_{11}^{\Trans}  \hat \mP_{k':k - 1}^{-1} \tilde \mL_{11}\big)^{-1}  \big(\tilde \mL_{21}^{\Trans} - \tilde \mL_{11}^{\Trans} \hat \mP_{k':k - 1}^{-1} \hat \mP_{k':k - 1, k}^{\UMV}\big), \label{eq:tilde Ak} \\
		\tilde \mL_{11} &= \begin{pmatrix}
			\mI_{(\ns - 1) \dx} & 0
		\end{pmatrix}
		\tilde \mL_k
		\begin{pmatrix}
			\mI_{(\ns - 1) \dd} & 0
		\end{pmatrix}^{\Trans}, 
	\tilde \mL_{21} = \begin{pmatrix}
		0 & \mI_{\dx} 
	\end{pmatrix}
	\tilde \mL_k
	\begin{pmatrix}
		\mI_{(\ns - 1) \dd} & 0
	\end{pmatrix}^{\Trans}. \label{eq:tilde Lk}
	\end{align}
\end{proposition}

\textit{Proof.}
The proof is similar to that of Theorem \ref{lem:CRLB computation}, and hence, omitted here.
$\hfill\square$

Note that the sizes of $\tilde \mL_{11}$ and $\tilde \mL_{21}$ given by \eqref{eq:tilde Lk} are independent of the time step $k$.
We next provide the relation between the PCRLB and the CRLB.

\begin{proposition} \label{prop:PCRLB}
	For the PCRLB given by \eqref{eq:PCRLB} and the CRLB given by \eqref{eq:CRLB:information inequality}, the following inequality holds:
	\begin{align}\label{eq:CRLB greater PCRLB}
		\trace\big(\B(\vd_{k - 1})\big) \geq \trace\big(\PB(\vd_{k - 1})\big).
	\end{align} 
\end{proposition}

\textit{Proof.}
	Denote 
$\mL_k^{\Trans} \hat \mP_{k' : k}^{-1} \mL_k = 
\begin{pmatrix}
\mA & \mB \\
\mB^{\Trans} & \mC
\end{pmatrix}$,
where $\mA \in \R^{(k - \ns) \dd \times (k - \ns) \dd}$, $\mB \in \R^{(k - \ns) \dd \times \ns \dd}$, $\mC \in \R^{\ns \dd \times \ns \dd}$.
	Then, we have
	\begin{align}
		\B(\vd_{k - 1}) 
		&=
		\begin{pmatrix}
			0& \mI_{\dd}	
		\end{pmatrix} \big(\mL_k^{\Trans} \hat \mP_{k' : k}^{-1} \mL_k\big)^{-1} \begin{pmatrix}
			0& \mI_{\dd}	
		\end{pmatrix}^{\Trans} \notag\\
		&= \begin{pmatrix}
			0& \mI_{\dd}	
		\end{pmatrix} \begin{pmatrix}
			0& \mI_{\ns \dd}	
		\end{pmatrix} \big(\mL_k^{\Trans} \hat \mP_{k' : k}^{-1} \mL_k\big)^{-1}  \begin{pmatrix}
			0& \mI_{\ns \dd}	
		\end{pmatrix}^{\Trans} \begin{pmatrix}
			0& \mI_{\dd}	
		\end{pmatrix}^{\Trans} \notag\\
		&= \begin{pmatrix}
			0& \mI_{\dd}	
		\end{pmatrix} (\mC^{-1} + \mC^{-1} \mB^{\Trans} (\mA - \mB \mC^{-1} \mB^{\Trans})^{-1} 
		\mB \mC^{-1})
		\begin{pmatrix}
			0& \mI_{\dd}	
		\end{pmatrix}^{\Trans}, \label{eq:CRLB:EX}\\
		\PB(\vd_{k - 1}) 
		&= \begin{pmatrix}
			0& \mI_{\dd}	
		\end{pmatrix} \Big(\begin{pmatrix}
			0 & \mI_{\ns \dd}
		\end{pmatrix} \mL_k^{\Trans} \hat \mP_{k' : k}^{-1} \mL_k 
		\begin{pmatrix}
			0 & \mI_{\ns \dd}	
		\end{pmatrix}^{\Trans} \Big)^{-1} 
		\begin{pmatrix}
			0& \mI_{\dd}	
		\end{pmatrix}^{\Trans}
		= \begin{pmatrix}
			0 & \mI_{\dd}	
		\end{pmatrix} \mC^{-1} \begin{pmatrix}
			0 & \mI_{\dd}	
		\end{pmatrix}^{\Trans}.\label{eq:PCRLB:EX}
	\end{align}
	Due to $\mC^{-1} \mB^{\Trans} (\mA - \mB \mC^{-1} \mB^{\Trans})^{-1} \mB \mC^{-1} \geq 0$, we obtain $\B(\vd_{k - 1}) \geq \PB(\vd_{k - 1})$, and thus, we have
	\begin{align*} 
		\trace\big(\B(\vd_{k - 1})\big) \geq \trace(\PB(\vd_{k - 1})),
	\end{align*}
which completes the proof.
$\hfill\square$

\begin{remark}
	From \eqref{eq:CRLB:EX} and \eqref{eq:PCRLB:EX}, we can directly derive the approximation error between $\PB(\vd_{k - 1})$ and $\B(\vd_{k - 1})$, denoted as
	\begin{align*}
		e=
	\begin{pmatrix}
		0 & \mI_{\dd}	
	\end{pmatrix} (\mC^{-1} \mB^{\Trans} (\mA - \mB \mC^{-1} \mB^{\Trans})^{-1} \mB \mC^{-1}) \begin{pmatrix}
	0 & \mI_{\dd}	
\end{pmatrix}^{\Trans}.
\end{align*}
\end{remark}

Based on Propositions \ref{prop:explicit:PCRLB} and  \ref{prop:PCRLB}, the original optimization problem \eqref{min:inject noise:0} can be relaxed as follows:
\begin{equation}
	\begin{aligned}\label{min:inject noise:1}
		\min_{\mSigma_k \in \sS} & \qquad \trace(\mSigma_k) \\
		\mathrm{s.t.} & \qquad \trace \Big(\big( \mG_{k-1}^{\Trans} (\mSigma_k + \tilde \mA_k)^{-1} \mG_{k-1}\big)^{-1}\Big) \geq \gamma, \
		\mSigma_k \geq \sigma \mI_{\dx},
	\end{aligned}
\end{equation}
It is not difficult to find that \eqref{min:inject noise:1} has the same form as \eqref{min:inject noise:0}. This indicates that we reduce the computational complexity without increasing the difficulty of solving the problem \eqref{min:inject noise:0}.

\textbf{\emph{3) Computational complexity.}}
Compared with the $\B(\vd_{k-1})$ given by \eqref{eq:explicit CRLB}, the computational complexity of the $\PB(\vd_{k - 1})$ given by \eqref{eq:explicit PCRLB} is greatly reduced, as specified by the following theorem.

\begin{theorem}[Computational complexity]\label{thm:computational complexity}
	From \eqref{eq:explicit CRLB} to \eqref{eq:explicit PCRLB}, the computational complexity reduces from $\mathcal{O}(k^3)$ to $\mathcal{O}(1)$.
\end{theorem}

\textit{Proof.}
	For calculating \eqref{eq:explicit CRLB}, the main computation lies in \eqref{eq:Ak} and is specified as follows. Specifically, for calculating 
	$\hat \mP_k^{\UMV} - \hat \mP_{k, k':k - 1}^{\UMV} \hat \mP_{k':k - 1}^{-1}  \hat \mP_{k':k - 1, k}^{\UMV}$, 
	the computational complexity is
$(\dx (\ns - 1))^2 + \dx^4 (\ns - 1)^2 + \dx^2$,
for calculating $\mL_{21}(k) -  \hat \mP_{k, k':k - 1}^{\UMV} \hat \mP_{k':k - 1}^{-1} \mL_{11}(k)$, the computational complexity is
$(\dx (\ns - 1))^3 + \dx^3 \dd (\ns - 1)^2 (k - 1) + \dx \dd (k - 1)$,
and for calculating $(\mL_{11}(k)^{\Trans}  \hat \mP_{k':k - 1}^{-1} \mL_{11}(k))^{-1}$, the computational complexity is
$(\dx (\ns - 1))^3 + \dx^2 \dd^2 (\ns - 1)^2 (k - 1)^2 + ((k - 1) \dd)^3$.
Additionally, for calculating 
\begin{align*}
(\mL_{21}(k) -  \hat \mP_{k, k':k - 1}^{\UMV} \hat \mP_{k':k - 1}^{-1} \mL_{11}(k))  (\mL_{11}(k)^{\Trans}  \hat \mP_{k':k - 1}^{-1} \mL_{11}(k))^{-1}  (\mL_{21}(k)^{\Trans} - \mL_{11}(k)^{\Trans} \hat \mP_{k':k - 1}^{-1} \hat \mP_{k':k - 1, k}^{\UMV}),
\end{align*}
the computational complexity is $\dx^2 \dd^2 (k - 1)^2$, and for calculating
$\hat \mP_k^{\UMV} - \hat \mP_{k, k':k - 1}^{\UMV} \hat \mP_{k':k - 1}^{-1}  \hat \mP_{k':k - 1, k}^{\UMV} + (\mL_{21}(k) - \hat \mP_{k, k':k - 1}^{\UMV} \hat \mP_{k':k - 1}^{-1} \mL_{11}(k))  (\mL_{11}(k)^{\Trans}  \hat \mP_{k':k - 1}^{-1} \mL_{11}(k))^{-1} (\mL_{21}(k)^{\Trans} - \mL_{11}(k)^{\Trans} \hat \mP_{k':k - 1}^{-1} \hat \mP_{k':k - 1, k}^{\UMV})$, the computational complexity is $\dx^2$. Hence, for calculating \eqref{eq:Ak}, the total computational complexity is 
	\begin{align}\label{eq:computational complexity for Ak}
		&\dd^3 (k - 1)^3 + \dx^2 \dd^2 (1 + (\ns - 1)^2 ) (k - 1)^2  + (\dx^3 \dd (\ns - 1)^2 + \dx \dd)(k - 1) \notag\\
		& + \dx^4 (\ns - 1)^2 + 3 \dx^3 (\ns - 1)^3 + 2 \dx^2.
	\end{align}
	For calculating \eqref{eq:explicit PCRLB}, the main computation lies in \eqref{eq:tilde Ak}. By replacing $k$ with $\ns$ in \eqref{eq:computational complexity for Ak}, we can obtain the computational complexity as follows:
	\begin{align}\label{eq:computational complexity for tilde Ak}
		&\dx^2 \dd^2 (\ns - 1)^4 + \dx^3 \dd (\ns - 1)^3 + \dd^3 (\ns - 1)^3  + 3 \dx^3 (\ns - 1)^3 + \dx^2 \dd^2 (\ns - 1)^2 + \dx^4 (\ns - 1)^2 \notag \\
		& + \dx \dd (\ns - 1) + 2 \dx^2.
	\end{align}
	From \eqref{eq:computational complexity for Ak} to \eqref{eq:computational complexity for tilde Ak}, the computational complexity reduces from $\mathcal{O}(k^3)$ to $\mathcal{O}(1)$. 
$\hfill\square$

Theorem \ref{thm:computational complexity}  indicates that from $\B(\vd_{k-1})$ to $\PB(\vd_{k - 1})$, the computational complexity is reduced from $\mathcal{O}(k^3)$ to $\mathcal{O}(1)$. In another word, the computational complexity of CRLB is reduced to be time-independent.

\textbf{\emph{4) Relaxed solution.}}
Since the analytic solution of the problem \eqref{min:inject noise:1} is hard to obtain,  we next provide a relaxed solution of \eqref{min:inject noise:1}.
Denote the singular value decomposition of $\mG_{k-1}$ by
\begin{align}\label{eq:L22 SVD}
	\mG_{k-1}= \mU_k 
	\begin{pmatrix}
		\mUpsilon_k & 0
	\end{pmatrix}^{\Trans}
	\mV_k,
\end{align}
where $\mU_k$ and $\mV_k$ are orthogonal matrices, $\mUpsilon_k \in \R^{\dd \times \dd}$ is a diagonal matrix, and denote
\begin{align}
	\mM_k = \mU_k^{\Trans} (\tilde \mA_k + \sigma \mI_{\dx}) \mU_k = \begin{pmatrix}
		\tilde \mA_{11} & \tilde \mA_{12}\\
		\tilde \mA_{21} & \tilde \mA_{22}
	\end{pmatrix}, \label{eq:Aij}
\end{align}
where $\tilde \mA_{11} \in \R^{\dd \times \dd}$. Then, the following theorem provides a relaxed solution of \eqref{min:inject noise:1}.

\begin{theorem}\label{thm:final relaxed solution}
	A relaxed solution of \eqref{min:inject noise:1} is given by
	\begin{align}\label{eq:Sigma_k^*}
		\mSigma_k =\mU_k \begin{pmatrix}
			\tilde \mSigma_{11}^* - \tilde \mA_{11} + \sigma \mI_{\dd} & 0\\
			0 & \sigma \mI_{\dx - \dd}
		\end{pmatrix} \mU_k^{\Trans},
	\end{align}
	where $\tilde \mSigma_{11}^*$ is the minimizer of the following semi-definite programming problem:
	\begin{equation}\label{min:final problem}
		\begin{aligned}
			\min_{\tilde \mSigma_{11} \in \sS} & \qquad \trace(\tilde \mSigma_{11}) \\
			\mathrm{s.t.} & \qquad \trace\big(\mUpsilon_k^{-2} (\tilde \mSigma_{11} - \tilde \mA_{12} \tilde \mA_{22}^{-1} \tilde \mA_{21})\big) \geq \gamma, \ \tilde \mSigma_{11} \geq \tilde \mA_{11}.
		\end{aligned}
	\end{equation}
\end{theorem}

\textit{Proof.}
	Problem \eqref{min:inject noise:1} is equivalent to
	\begin{equation}\label{min:inject noise:2}
		\begin{aligned}
			\min_{\mSigma_k \in \sS} & \qquad \trace(\mSigma_k + \tilde \mA_k) \\
			\mathrm{s.t.} & \qquad \trace \Big(\big( \mG_{k-1}^{\Trans} (\mSigma_k + \tilde \mA_k)^{-1} \mG_{k-1}\big)^{-1}\Big) \geq \gamma, \ 
			\mSigma_k + \tilde \mA_k \geq \tilde \mA_k + \sigma \mI_{\dx}.
		\end{aligned}
	\end{equation}
	From \eqref{eq:L22 SVD}, we have
$
       \mG_{k-1}^{\Trans} (\mSigma_k + \tilde \mA_k)^{-1} \mG_{k-1}
		=  \mV_k^{\Trans} \begin{pmatrix} \mUpsilon_k & 0 \end{pmatrix} \mU_k^{\Trans} (\mSigma_k + \tilde \mA_k)^{-1}  \mU_k \begin{pmatrix} \mUpsilon_k & 0 \end{pmatrix}^{\Trans} \mV_k.
$
	Denote
	$
		\tilde \mSigma_k^{-1} = \mU_k^{\Trans} (\mSigma_k + \tilde \mA_k)^{-1} \mU_k = \begin{pmatrix}
			\tilde \mSigma_{11} & \tilde \mSigma_{12}\\
			\tilde \mSigma_{21} & \tilde \mSigma_{22}
		\end{pmatrix}^{-1}
	$
	with $\tilde \mSigma_{11} \in \R^{\dd \times \dd}$.
	Then, the problem \eqref{min:inject noise:2} is equivalent to
	\begin{equation}\label{min:inject noise:3}
		\begin{aligned}
			\min_{\tilde \mSigma_k \in \sS} & \qquad \trace(\tilde \mSigma_k) \\
			\mathrm{s.t.} & \qquad \trace \bigg( \Big( \mV_k^{\Trans} \begin{pmatrix} \mUpsilon_k & 0 \end{pmatrix}  \tilde \mSigma_k^{-1} \begin{pmatrix} \mUpsilon_k & 0 \end{pmatrix}^{\Trans}  \mV_k \Big)^{-1} \bigg) 
			\geq \gamma, \ 
			 \tilde \mSigma_k \geq \mU_k^{\Trans} (\tilde \mA_k + \sigma \mI_{\dx}) \mU_k.
		\end{aligned}
	\end{equation}
	Here, we use the fact that
$
		\trace (\mSigma_k + \tilde \mA_k) = \trace \big(\mU_k^{\Trans} (\mSigma_k + \tilde \mA_k) \mU_k\big) = \trace (\tilde \mSigma_k).
$
	Denote
	$\tilde \mSigma_k^{-1} = \begin{pmatrix}
			\mS_{11} & \mS_{12}\\
			\mS_{21} & \mS_{22}
		\end{pmatrix}$.
	Then, we have
$\mS_{11} = (\tilde \mSigma_{11} - \tilde \mSigma_{12} \tilde \mSigma_{22}^{-1} \tilde \mSigma_{21})^{-1} \in \R^{\dd \times \dd}$,
and
	\begin{align*}
		\trace \bigg( \Big( \mV_k^{\Trans} \begin{pmatrix} \mUpsilon_k & 0 \end{pmatrix} \tilde \mSigma_k^{-1} \begin{pmatrix} \mUpsilon_k & 0 \end{pmatrix}^{\Trans}  \mV_k \Big)^{-1} \bigg)
		= \trace\big((\mUpsilon_k \mS_{11} \mUpsilon_k)^{-1}\big) 
		= \trace\big(\mUpsilon_k^{-2} (\tilde \mSigma_{11} - \tilde \mSigma_{12} \tilde \mSigma_{22}^{-1} \tilde \mSigma_{21})\big).
	\end{align*}
	Thus, the problem \eqref{min:inject noise:3} is equivalent to
	\begin{equation}\label{min:inject noise:4}
		\begin{aligned}
			\min_{\tilde \mSigma_{11}, \tilde \mSigma_{22} \in  \sS} & \qquad \trace(\tilde \mSigma_{11}) + \trace(\tilde \mSigma_{22}) \\
			\mathrm{s.t.} & \qquad \trace\big(\mUpsilon_k^{-2} (\tilde \mSigma_{11} - \tilde \mSigma_{12} \tilde \mSigma_{22}^{-1} \tilde \mSigma_{21})\big) \geq \gamma, \ \begin{pmatrix}
				\tilde \mSigma_{11} & \tilde \mSigma_{12}\\
				\tilde \mSigma_{21} & \tilde \mSigma_{22}
			\end{pmatrix} \geq \begin{pmatrix}
				\tilde \mA_{11} & \tilde \mA_{12}\\
				\tilde \mA_{21} & \tilde \mA_{22}
			\end{pmatrix}.
		\end{aligned}
	\end{equation}
	By letting $\tilde \mSigma_{22} = \tilde \mA_{22}$, the problem \eqref{min:inject noise:4} can be relaxed to
	\begin{equation}
		\begin{aligned}\label{min:relaxed problem}
			\min_{\tilde \mSigma_{11} \in \sS} & \qquad \trace(\tilde \mSigma_{11}) \\
			\mathrm{s.t.} & \qquad \trace\big(\mUpsilon_k^{-2} (\tilde \mSigma_{11} - \tilde \mSigma_{12} \tilde \mA_{22}^{-1} \tilde \mSigma_{21})\big) \geq \gamma, \  \begin{pmatrix}
				\tilde \mSigma_{11} & \tilde \mSigma_{12}\\
				\tilde \mSigma_{21} & \tilde \mA_{22}
			\end{pmatrix} \geq \begin{pmatrix}
				\tilde \mA_{11} & \tilde \mA_{12}\\
				\tilde \mA_{21} & \tilde \mA_{22}
			\end{pmatrix}.
		\end{aligned}
	\end{equation}
	We adopt this relaxation because
$\begin{pmatrix}
			\tilde \mSigma_{11} & \tilde \mSigma_{12}\\
			\tilde \mSigma_{21} & \tilde \mSigma_{22}
		\end{pmatrix} \geq
		\begin{pmatrix}
			\tilde \mA_{11} & \tilde \mA_{12}\\
			\tilde \mA_{21} & \tilde \mA_{22}
		\end{pmatrix} \geq 0$
	implies
	$\tilde \mSigma_{11} \geq \tilde \mA_{11}$ and $\tilde \mSigma_{22} \geq \tilde \mA_{22}$. 
	From the sufficient and necessary condition for the positive semi-definiteness of a matrix in terms of a generalized Schur complement \cite{Zhang2005The}, we have
	\begin{align*}
		&\begin{pmatrix}
			\tilde \mSigma_{11} - \tilde \mA_{11} & \tilde \mSigma_{12} - \tilde \mA_{12} \\
			\tilde \mSigma_{21} - \tilde \mA_{21} & 0
		\end{pmatrix} \geq 0
		\iff \tilde \mSigma_{11} - \tilde \mA_{11}  \geq 0,\  \tilde \mSigma_{21} - \tilde \mA_{21} = \tilde \mSigma_{12} - \tilde \mA_{12} = 0.
	\end{align*}
	Thus, the problem \eqref{min:relaxed problem} is equivalent to the problem \eqref{min:final problem}. 
	Further, let $\mSigma_{11}^*$ be the solution of \eqref{min:final problem}. Then, 
	\begin{align*}
		\mSigma_k &= \mU_k \tilde \mSigma_k \mU_k^{\Trans} - \tilde \mA_k \\
		&= \mU_k \begin{pmatrix}
			\tilde \mSigma_{11}^* & \tilde \mA_{12} \\
			\tilde \mA_{21}     & \tilde \mA_{22}
		\end{pmatrix} \mU_k^{\Trans} - \tilde \mA_k \\
		&= \mU_k \begin{pmatrix}
			\tilde \mA_{11} & \tilde \mA_{12} \\
			\tilde \mA_{21}     & \tilde \mA_{22}
		\end{pmatrix} \mU_k^{\Trans} - \tilde \mA_k  + \mU_k \begin{pmatrix}
			\tilde \mSigma_{11}^* - \tilde \mA_{11} & 0 \\
			0 & 0
		\end{pmatrix} \mU_k^{\Trans} \\
		&= \sigma \mI_{\dx} + \mU_k \begin{pmatrix}
			\tilde \mSigma_{11}^* - \tilde \mA_{11} & 0 \\
			0 & 0
		\end{pmatrix} \mU_k^{\Trans} \\
		&=\mU_k \begin{pmatrix}
			\tilde \mSigma_{11}^* - \tilde \mA_{11} + \sigma \mI_{\dd} & 0\\
			0 & \sigma \mI_{\dx - \dd}
		\end{pmatrix} \mU_k^{\Trans}.
	\end{align*} 
This completes the proof. 
$\hfill\square$

Theorem \ref{thm:final relaxed solution} indicates that the problem of determining the covariance matrix $\mSigma_k$ of the perturbed noise is finally converted to the problem of solving the semi-definite programming problem given by \eqref{min:final problem}. 
Instead of solving the original optimization problem \eqref{min:inject noise:0} directly, we relax it to the semi-definite programming problem  \eqref{min:final problem}, which has the following benefits:
i) the computational complexity is greatly reduced;
ii) the semi-definite programming problem can be solved efficiently; 
iii) the privacy level remains no less than $\gamma$. 

\textbf{\emph{5) Algorithm.}}
The proposed privacy-preserving state estimation algorithm with low complexity is summarized in Algorithm \ref{alg}.  
The semi-definite programming problem \eqref{min:final problem} therein can be efficiently solved by, e.g., the CVX package. For more details about the CVX package, the readers are referred to \cite{cvx}.

\begin{algorithm}[htbp]\label{alg}
	\renewcommand{\algorithmicrequire}{\textbf{Input:}}
	\renewcommand{\algorithmicensure}{\textbf{Output:}}
	\caption{Privacy-preserving state estimation algorithm with low complexity}
	\label{alg}
	\begin{algorithmic}[1]
		\REQUIRE $\hat \vx_{k - 1}^{\UMV}$, $\hat \mS_{k - 1}^{\UMV}$, $\ns$, $\sigma$, $\gamma$
		\STATE \emph{\textbf{Prediction:}}
		\STATE Calculate $\hat \vx_k^-$ and $\hat \mS_k^-$ using \eqref{eq:UMV:pre:mean} and \eqref{eq:UMV:pre:cov}.
		\STATE \emph{\textbf{Update:}}
		\STATE Calculate $\hat \vx_k^{\UMV}$ and $\hat \mS_k^{\UMV}$ based on  $\hat \vx_{k - 1}^{\UMV}$ and $\hat \mS_{k - 1}^{\UMV}$ using \eqref{eq:UMV estimate} and \eqref{eq:UMV:update:cov}.
		\STATE \emph{\textbf{Calculate the covariance matrix of the perturbed noise:}}
		\STATE Set $k' = k - \ns + 1$.
		\STATE Calculate each sub-block of $\hat \mP_{k' : k}^{\UMV}$ recursively using \eqref{eq:cov:xk and xk}--\eqref{eq:cov:hat xk and xk+j hat}, and then obtain $\hat \mP_{k, k':k - 1}^{\UMV}$ and $\hat \mP_{k':k - 1, k}^{\UMV}$ using \eqref{eq:hat P:UMV:aug}.
		\STATE Calculate $\hat \mP_{k':k - 1}^{-1}$ using \eqref{eq:hat:P:aug}.
		\STATE Calculate $\tilde \mL_k$ using \eqref{eq:explicit:tilde Lk}, and then obtain $\tilde \mL_{11}$ and $\tilde \mL_{21}$ using \eqref{eq:tilde Lk}.
		\STATE Calculate $\tilde \mA_k$ using \eqref{eq:tilde Ak}.
		\STATE Perform singular value decomposition on $\mG_{k-1}$ to obtain $\mUpsilon_k$ and $\mU_k$ using \eqref{eq:L22 SVD}.
		\STATE Calculate $\mM_k$ using \eqref{eq:Aij}  with the pre-set $\sigma$ 
		to obtain
		$\tilde \mA_{11}$, $\tilde \mA_{12}$, $\tilde \mA_{21}$, and $\tilde \mA_{22}$.
		\STATE Solve the semi-definite programming problem \eqref{min:final problem} to obtain $\tilde \mSigma_{11}^*$, with $\gamma$ being the pre-set privacy level.
		\STATE Calculate $\mSigma_k$ using \eqref{eq:Sigma_k^*}.
		\STATE \emph{\textbf{Privacy-preserving state estimation:}}
		\STATE Generate $\valpha_k \sim \N(0, \mSigma_k)$.
		\STATE Set $\hat \vx_{k} = \hat \vx_k^{\UMV} + \valpha_k$.
		\STATE Set $\hat \mS_k = \hat \mS_k^{\UMV} + \mSigma_k$.
		\ENSURE $\hat \vx_k$, $\hat \mS_k$
	\end{algorithmic}
\end{algorithm}


\subsection{Relation to differential privacy}

This subsection shows that Algorithm \ref{alg} also ensures $(\epsilon, \delta)$-differential privacy. To this end, we first define the sensitivity of Algorithm \ref{alg} as follows, which determines how much perturbed noise should be added.

\begin{definition}[Sensitivity]
	Suppose that $\mathbb R^{\dd}$ is equipped with an adjacency relation $\adj$. The sensitivity of a query $q: \mathbb R^{\dd} \rightarrow \mathbb R^{n}$ is defined as
	\begin{align*}
		\Delta_{\mA}q \coloneqq \sup_{\vd_k, \vd_k': \adj(\vd_k, \vd_k')} \|q(\vd_k) - q(\vd_k') \|_{\mA}, \ \mA > 0.
	\end{align*}
\end{definition}

At time step $k$, the change in $\vd_{k-1}$ only affects $\hat \vx_{k}$ rather than $\hat \vx_{k' : k - 1}$, and thus, the Gaussian mechanism $\M_q: \mathbb R^{\dx} \times \mathbb R^{\dx} \rightarrow \mathbb R^{\dx}$ is defined by $\M_q(\vd_{k-1}) = \hat \vx_k = q(\vd_{k-1}) + \omega$, where
$\omega \sim \N(0, \hat \mP_k)$, $\hat \mP_k = \hat \mP_k^{\UMV} + \mSigma_k$ is the covariance matrix of $\hat \vx_k$, and $q(\vd_k) =\E [\hat \vx_k] = \mK_k \mH_k \mG_{k - 1} \vd_{k-1} + \vc_k$ with $\vc_k$ being a constant vector.  
Based on these analyses, we can show exactly what level of differential privacy  Algorithm \ref{alg} can ensure in the following theorem.

\begin{theorem}[Differential privacy]\label{thm:DP}
	For any $\epsilon \geq 0$, Algorithm \ref{alg} is $(\epsilon, \delta)$-differentially private with $\delta = \mathcal{Q}(\xi) = \frac{1}{\sqrt{2 \pi}} \int_{\xi}^{\infty} \exp\{-\frac{z^2}{2}\}\diff z$ and
	\begin{align}\label{eq:xi}
		\xi =  -\frac{\Delta_{\hat \mP_k^{-1}} q}{2} + \frac{\epsilon}{\Delta_{\hat \mP_k^{-1}} q}.
	\end{align}
\end{theorem}

\textit{Proof.}
	Let $\vd_{k-1}$, $\vd_{k'-1}$ be two adjacent elements in $\mathbb R^{\dd}$, and denote $\vv \coloneqq \mK_k \mH_k \mG_{k - 1} (\vd_{k-1} - \vd_{k'-1})$. Then, for any Borel set $S \in \mathbb R^{\dx}$, we have
	\begin{align*}
		&P(\M_q(\vd_{k-1}) \in S) \\
		&= \int_S \N(\vu; q(\vd_{k-1}), \hat \mP_k) \diff \vu \\
		&= \int_S (2 \pi)^{-\frac{\dx}{2}} \det(\hat \mP_k)^{-\frac{1}{2}} \exp\bigg\{-\frac{1}{2} \big\| \vu - q(\vd_{k-1}) \big\|^2_{\hat \mP_k^{-1}}\bigg\} \diff \vu \\
		&= \int_S (2 \pi)^{-\frac{\dx}{2}} \det(\hat \mP_k)^{-\frac{1}{2}} \exp\bigg\{-\frac{1}{2} \big\| \vu - q(\vd_{k'-1}) \big\|^2_{\hat \mP_k^{-1}}\bigg\}  \exp\bigg\{ \big(\vu - q(\vd_{k'-1})\big)^{\Trans} \hat \mP_k^{-1} \vv - \frac{1}{2} \| \vv \|^2_{\hat \mP_k^{-1}} \bigg\} \diff \vu.
	\end{align*} 
	Let $f(\vu) = (\vu - q(\vd_{k'-1}))^{\Trans} \hat \mP_k^{-1} \vv - \frac{1}{2} \| \vv \|^2_{\hat \mP_k^{-1}}$, $A = \{ \vu | f(\vu) \leq \epsilon \}$. Then, 
	\begin{align*}
		&P(\M_q(\vd_{k-1}) \in S) \\
		&= \int_{S \cap A} (2 \pi)^{-\frac{\dx}{2}} \det(\hat \mP_k)^{-\frac{1}{2}} \exp\bigg\{-\frac{1}{2} \big\| \vu - q(\vd_{k'-1}) \big\|^2_{\hat \mP_k^{-1}}\bigg\}  \exp\{ f(\vu) \} \diff \vu + \int_{S \cap {A^c}} \N(\vu; q(\vd_{k-1}), \hat \mP_k) \diff \vu \\
		&\leq e^{\epsilon} P(\M_q(\vd_{k'-1}) \in S) + \int_S \N(\vu; q(\vd_{k-1}), \hat \mP_k) \mathcal{I}_{[f(\vu) > \epsilon]} \diff \vu,
	\end{align*}
	where $A^c$ is the complement set to $A$, and $\mathcal{I}_{[f(\vu) > \epsilon]}$ is an indicative function defined as
$
		\mathcal{I}_{[f(\vu) > \epsilon]} = \begin{cases}
			1 & f(\vu) > \epsilon \\
			0 & f(\vu) \leq \epsilon
		\end{cases}
$. 
	Let $\vy = \hat \mP_k^{-\frac{1}{2}} (\vu - q(\vd_{k-1}))$. Then, we have
	\begin{align*}
		P(\M_q(\vd_{k-1}) \in S) 
		&\leq  e^{\epsilon} P(\M_q(\vd_{k'-1}) \in S)  + \int_S \N(\vy; 0, \mI_{\dx}) \mathcal{I}_{[\vv^{\Trans} \hat \mP_k^{-\frac{1}{2}} \vy > -\frac{1}{2} \| \vv \|^2_{\hat \mP_k^{-1}} + \epsilon]} \diff \vy \\
		&\leq e^{\epsilon} P(\M_q(\vd_{k'-1}) \in S) + \mathcal{Q}\bigg(-\frac{1}{2} \| \vv \|_{\hat \mP_k^{-1}} + \frac{\epsilon}{\| \vv \|_{\hat \mP_k^{-1}}}\bigg) \\
		&\leq e^{\epsilon} P(\M_q(\vd_{k'-1}) \in S) + \mathcal{Q}\bigg( -\frac{\Delta_{\hat \mP_k^{-1}} q}{2} + \frac{\epsilon}{\Delta_{\hat \mP_k^{-1}} q} \bigg).
	\end{align*}
	Thus, for any $\epsilon \geq 0$, $\M_q$ is $(\epsilon, \mathcal{Q}(\xi))$-differentially private, where $\xi$ is given by \eqref{eq:xi}. 
$\hfill\square$

\begin{remark}
	As established by Theorem  \ref{thm:DP}, for any fixed $\epsilon \geq 0$, a larger value  $\gamma$ leads to a smaller value of $\delta$. This relationship can be further detailed: 
	increasing $\gamma$ results in a larger matrix $\hat \mP_k$. Furthermore, it follows from \eqref{eq:xi} that a larger $\hat \mP_k$ implies a greater $\xi$, which consequently leads to a smaller $\delta$. 
	This relationship is further illustrated in Fig. \ref{fig:DP}.
\end{remark}



\section{Examples}\label{sec:simulation}

In this section, we demonstrate the effectiveness of the proposed algorithm through two examples.

\subsection{Practical case: Building occupancy}

\begin{figure}[htbp]
	\centering
	\subfloat[CO$_2$ level and its estimates by unbiased minimum-variance and privacy-preserving state estimates.]{
		\includegraphics[width=0.6\textwidth]{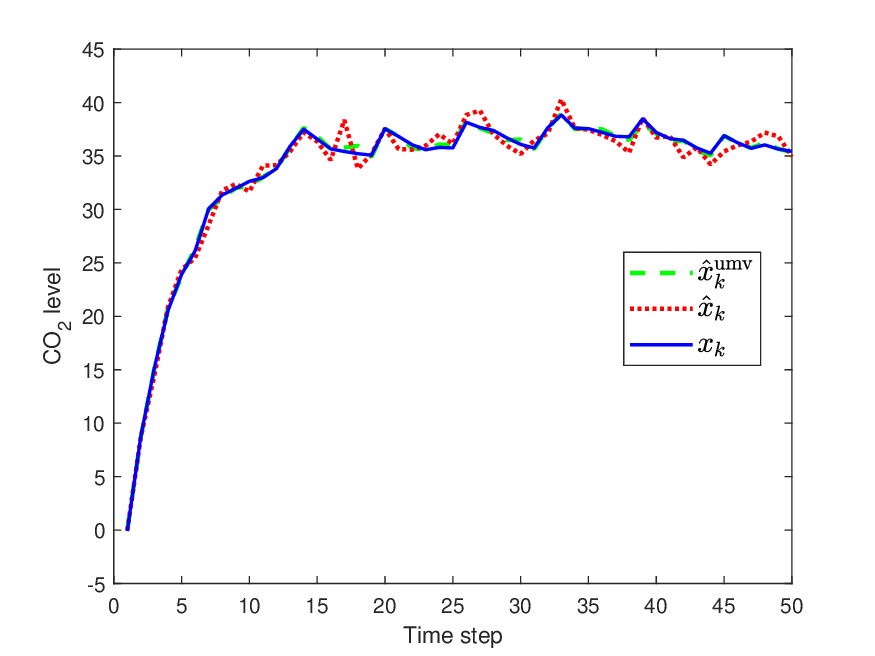}
		\label{fig:state estimate1}
	}
	\hfill
	\subfloat[Occupancy and its estimates by the adversary using unbiased minimum-variance and privacy-preserving state estimates.]{
		\includegraphics[width=0.6\textwidth]{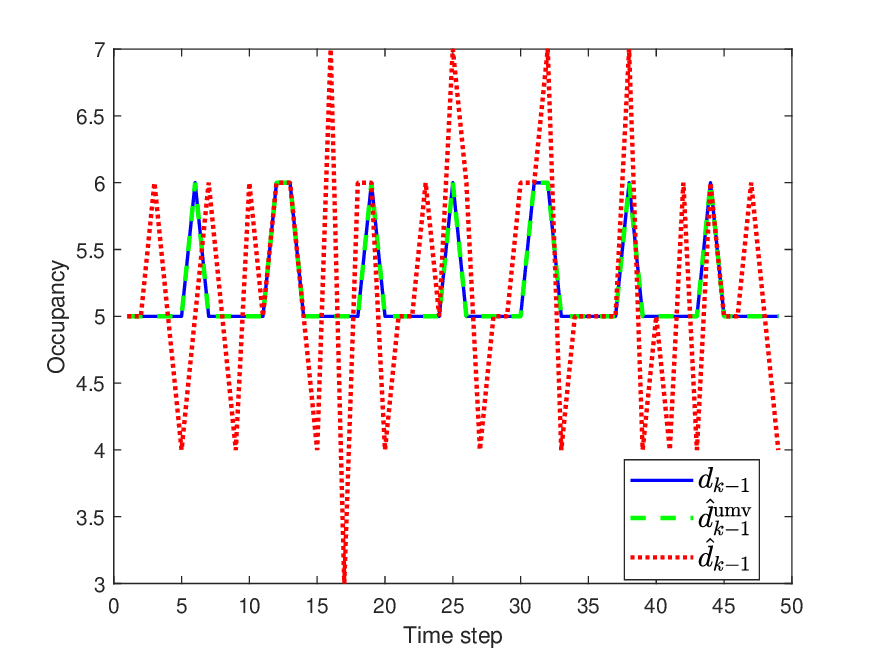}
		\label{fig:input estimate1}
	}
	\caption{CO$_2$ level, occupancy and their estimates.}
	\label{fig:2222}
\end{figure}

Consider Example \ref{subsec:motivate:exam} and set
$\sigma = 10^{-4}$, $\ns = 2$ and $\gamma = 0.5$ in Algorithm \ref{alg}. 
Figs. \ref{fig:state estimate1} and \ref{fig:input estimate1} highlight that the estimated trajectory of CO$_2$ level by the proposed algorithm is very close to the real one. In contrast, the estimated trajectory of the occupancy is not identical with the real one almost every time step. This demonstrates that the proposed algorithm performs well in estimating CO$_2$ level and protecting occupancy, ensuring privacy and utility simultaneously.

%

\subsection{Numerical case: A two-dimensional model}\label{sec:example:2}

%

Consider the two-dimensional dynamic system
$
	\vx_{k + 1} = \mF \vx_k + \mG d_k + \vw_k,
$
where $\vx_k \in \R^2$, $d_k \in \R$, $\mF = \begin{pmatrix}
	1 & 1 \\
	0 & 1
\end{pmatrix}$, $\mG = \begin{pmatrix}
	0.5 &
	0.5
\end{pmatrix}^{\Trans}$, $\vw_k \sim \N(0, \mQ_k)$ with $\mQ_k = 2 \mI_2$. The initial state $\vx_0$ was drawn from the Gaussian distribution $\N(\bar \vx_0, \mP_0)$ with
$\bar \vx_0 = \begin{pmatrix}
	2 & 2
\end{pmatrix}^{\Trans}$ and $\mP_0 = 0.1 \mI_2$. The exogenous input $d_k$ is generated independently and identically distributed from a uniform distribution over the interval $[0, 5]$. 
The measurement equation is
$
	\vy_k = \mH \vx_k + \vv_k,
$
where $\vy_k \in \R^2$, $\mH = \mI_2$, $\vv_k \sim \N(0, \mR_k)$ with $\mR_k = \mI_2$. We set $\gamma = 11$, $\ns = 3$ and $\sigma = 10^{-4}$ in Algorithm \ref{alg}. In addition to the PCRLB derived in \eqref{eq:explicit PCRLB}, we also employ the CRLB presented in \eqref{eq:explicit CRLB} for comparative analysis.

\begin{figure}[htbp]
	\centering
	\subfloat[MSEs of unbiased minimum-variance and privacy-preserving state estimates.]{
		\includegraphics[width=0.6\textwidth]{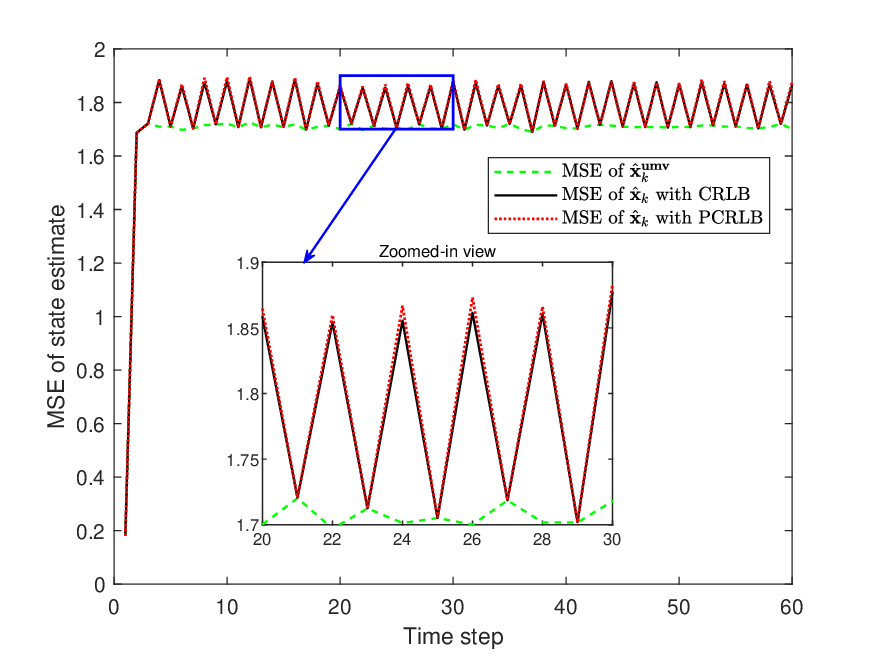}
		\label{fig:state MSE}
	}
	\hfill
	\subfloat[Trajectories of real states, unbiased minimum-variance state estimates, and privacy-preserving state estimates.]{
		\includegraphics[width=0.6\textwidth]{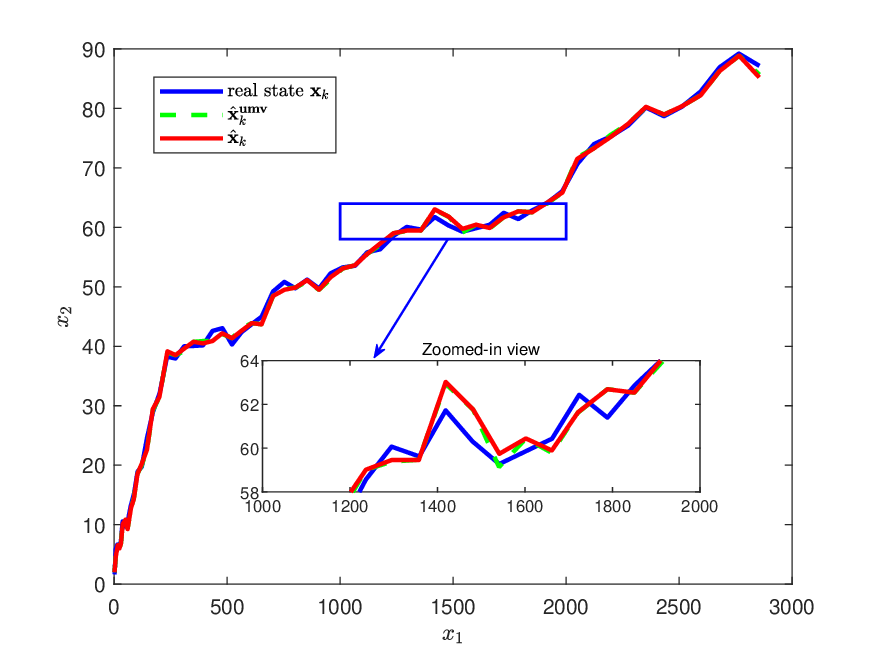}
		\label{fig:StateTra}
	}
	\caption{Comparison of unbiased minimum-variance and privacy-preserving state estimates.}
	\label{fig:3333}
\end{figure}

Fig. \ref{fig:state MSE} reports the MSEs of the proposed algorithm over 50 time steps and 500 Monte Carlo runs (i.e., the black solid line and the red dotted line), where the unbiased minimum-variance state estimate serves as a benchmark (i.e., the green dashed line). 
In Fig. \ref{fig:state MSE}, both the black solid line and the red dotted line lie above the green dashed line, which revels that the MSEs of the proposed algorithm are greater than those of the unbiased minimum-variance state estimate. This is reasonable due to the perturbed noise. 
Besides, by noting that the red dotted line and the black solid line exhibit highly similar estimation performance, we can infer that the approximation errors of PCRLB and CRLB are negligible in this example, which is specifically reflected in their nearly identical accuracy in state estimation.
Fig. \ref{fig:StateTra} depicts the real state trajectory and the estimated trajectories by the unbiased minimum-variance state estimator and the proposed algorithm. This figure reveals that the proposed algorithm has good estimation performance similar to the that of the unbiased minimum-variance estimator.

\begin{figure}[htbp]
	\centering
	\includegraphics[width=0.6\textwidth]{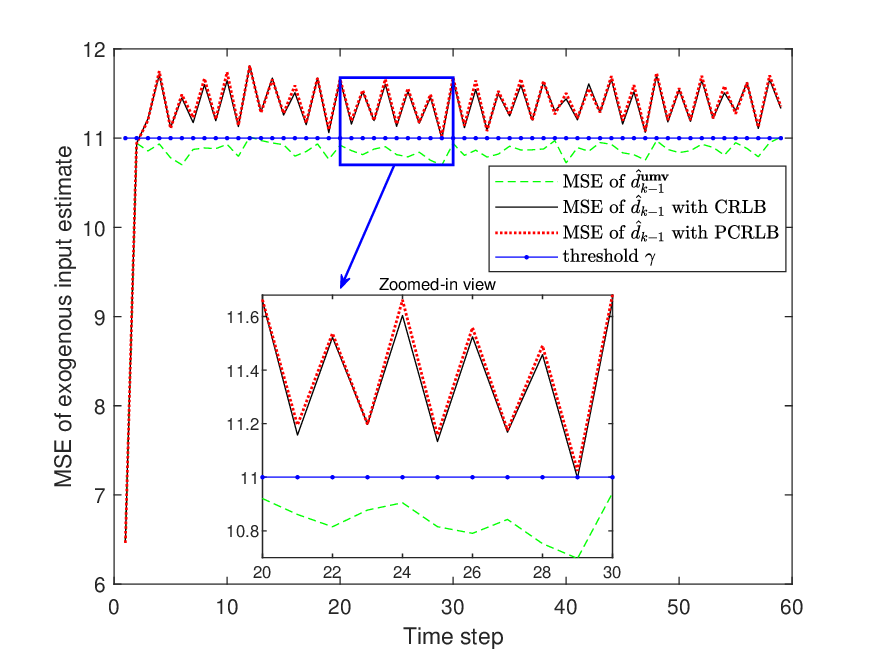}
	\caption{MSEs of adversary's estimates for $d_k$ with unbiased minimum-variance and privacy-preserving state estimates.}
	\label{fig:input MSE}
\end{figure}

Fig. \ref{fig:input MSE} highlights that the MSEs of the adversary's estimates for $d_k$ with either unbiased minimum-variance (green dashed line) or privacy-preserving state estimates (black solid and red dotted lines).  
We have the following two observations: 
i) MSEs of $\hat d_k$ with privacy-preserving state estimates are greater than those of $\hat d_k^{\UMV}$ with unbiased minimum-variance state estimates. This is resulted from the perturbed noise strategy.
ii) MSEs of $\hat d_k$ with privacy-preserving state estimates are greater than the threshold $\gamma$, while MSEs of $\hat d_k^{\UMV}$ with unbiased minimum-variance state estimates are smaller than the threshold $\gamma$. This phenomenon indicates that the proposed privacy-preserving state estimates do protect the exogenous input such that the MSEs of the adversary's estimates for $d_k$ are not less than $\gamma$.

\begin{figure}[htbp]
	\centering
	\includegraphics[width=0.6\textwidth]{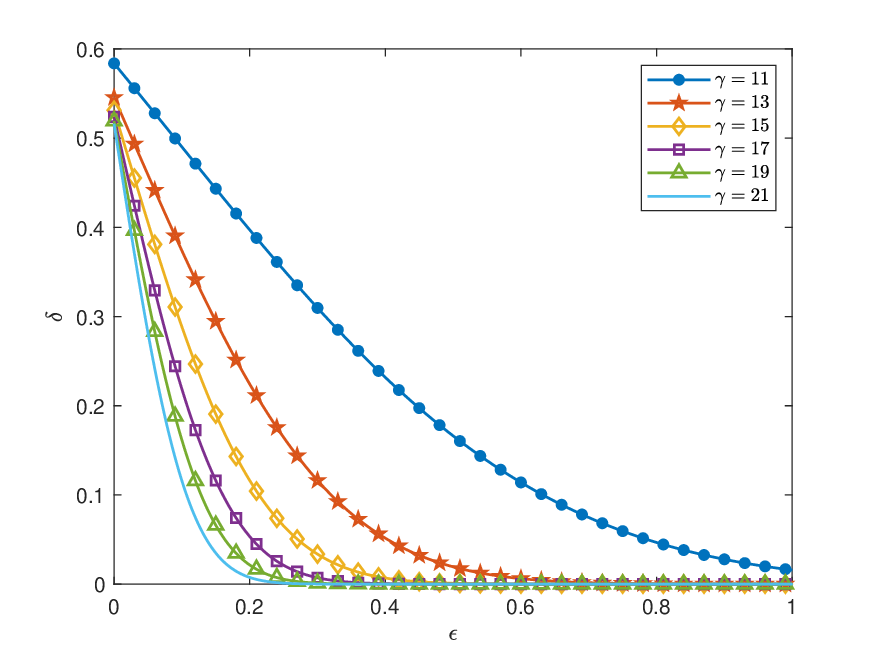}
	\caption{Correlation between the proposed CRLB-based method and differential privacy.}
	\label{fig:DP}
\end{figure}

Fig. \ref{fig:DP} quantitatively displays the correlation between the proposed CRLB-based method and differential privacy. We can see that as the parameter $\gamma$ measuring the privacy level increases, the curves approach the origin more closely, indicating that the level of differential privacy is higher.

\begin{table}[htbp]
	\centering
	\caption{Averaged MSEs of $\hat \vx$ and $\hat d$ under different privacy level $\gamma$.}
	\label{fig: private state MSE}
	\begin{tabular}{c|c|c|c|c|c}
		\hline $\gamma$              & 9       & 10      & 11      & 12     & 13 \\
		\hline $\MSE$ of $\hat \vx$ & 1.6852   & 1.6852  & 1.7664
		& 1.9998 & 2.2488 \\
		\hline $\MSE$ of $\hat d$ & 10.7997 & 10.7997 & 11.3009 & 12.8472 & 14.4323\\
		\hline
	\end{tabular}
\end{table}

Table \ref{fig: private state MSE} reports the MSEs of $\hat \vx$ and $\hat d$ averaged over 50 time steps and 500 Monte Carlo runs under different privacy level $\gamma$. It is observed that as $\gamma$ increases, both the MSE of $\hat \vx$ and the MSE of the adversary's estimate for $d_k$ increase. This is because, as the privacy level increases, the accuracy of state estimation decreases while the level of exogenous input protection strengthens, demonstrating the trade-off between the state estimation accuracy and the privacy level. 
Additionally, it is noteworthy that when $\gamma = 9$ and $\gamma = 10$, the MSE values are identical. This phenomenon stems from the fact that, under the framework of unbiased minimum variance state estimation,  the MSE associated with the adversary's optimal unbiased estimate of $d$ exceeds $10$, thereby eliminating the necessity for additional noise injection.

\section{Conclusion}\label{sec:conclusion}

We have developed a CRLB-based privacy-preserving state estimation algorithm with low complexity, which also ensures $(\epsilon, \delta)$-differential privacy.
Specifically, 
by perturbing the unbiased minimum-variance state estimate with a zero-mean Gaussian noise, we have designed a noisy state estimate that prevents the adversary from inferring the exogenous inputs. 
Adopting the CRLB allows constraining the MSE of the adversary's estimate for the exogenous inputs.
By minimizing the MSE of the noisy state estimate subject to a certain privacy level measured by CRLB, we have ensured privacy and utility by solving a non-convex constrained optimization. 
Additionally, we have provided explicit and low-complexity calculations for CRLB, significantly reducing the computational complexity from $\mathcal{O}(k^3)$ to $\mathcal{O}(1)$.
Furthermore, we have solved the constrained optimization efficiently by providing a relaxed solution.
Finally, we have demonstrated the effectiveness of the proposed algorithm through two examples, including a practical scenario for protecting building occupancy.

In practice, the measurements are
usually collected by some sensors in a network structure, so our future work includes studying the privacy-preserving state estimation problem for multi-sensor systems.

\bibliographystyle{IEEEtran}
\bibliography{mybibfile}



\end{document}